\documentclass[fleqn,usenatbib]{mnras}

\usepackage{newtxtext,newtxmath}

\usepackage[T1]{fontenc}
\usepackage[utf8]{inputenc}

\newcommand*{\bba}{$^{\scriptstyle 3\mathrm{D}}$B{\sc {arolo}}}

\DeclareRobustCommand{\VAN}[3]{#2}
\let\VANthebibliography\thebibliography
\def\thebibliography{\DeclareRobustCommand{\VAN}[3]{##3}\VANthebibliography}

\usepackage{graphicx}	
\usepackage{amsmath}	

\title[The assembly and fate of a giant disc galaxy]{The assembly and fate of a giant disc galaxy in a protocluster at $z = 3$}

\author[F. Rizzo et al.]{
Francesca Rizzo,$^{1}$\thanks{E-mail: rizzo@astro.rug.nl}
Pavel E. Mancera Piña,$^{2}$
Gabriele Pezzulli,$^{1}$
Giulia Despali$^{3, 4, 5}$
\\
$^{1}$Kapteyn Astronomical Institute, University of Groningen, Landleven 12, 9747 AD, Groningen, The Netherlands\\
$^{2}$Leiden Observatory, Leiden University, P.O. Box 9513, 2300 RA, Leiden, The Netherlands\\
$^{3}$ Dipartimento di Fisica e Astronomia "Augusto Righi", Alma Mater Studiorum Università di Bologna, via Gobetti 93/2, I-40129 Bologna, Italy\\
$^{4}$ INAF-Osservatorio di Astrofisica e Scienza dello Spazio di Bologna, Via Piero Gobetti 93/3, I-40129 Bologna, Italy\\
$^{5}$ INFN-Sezione di Bologna, Viale Berti Pichat 6/2, I-40127 Bologna, Italy\\
}

\date{Accepted XXX. Received YYY; in original form ZZZ}

\pubyear{\the\year{}}

\begin{document}
\label{firstpage}
\pagerange{\pageref{firstpage}--\pageref{lastpage}}
\maketitle

\begin{abstract}

Recent \textit{JWST} observations revealed two massive ($M_{\star} \gtrsim 10^{11}\,M_{\odot}$), unexpectedly large spiral galaxies at $z \sim 3$, both in overdense environments. We focus on one of these, ADF22.1 at $z = 3.09$, which hosts an active galactic nucleus (AGN), exploiting its extended [CII] emission ($\sim$30~kpc in diameter) with high-resolution observations from the Atacama Large Millimetre Array and \textit{JWST}. 
We find a flat outer rotation curve reaching $\sim$530~km\,s$^{-1}$, and perform, for the first time 
for a system of this type, a rotation-curve decomposition. 
We infer a dark-matter halo mass of $\log(M_{200}/M_{\odot})=12.9^{+0.4}_{-0.3}$, a baryon-to-halo mass ratio of $0.4^{+0.6}_{-0.3}$ in units of the cosmological baryon fraction, and a ratio between the baryonic and dark-matter halo specific angular momentum of $1.0^{+0.7}_{-0.5}$. Comparing these quantities with those of local galaxies, we find that ADF22.1 is indistinguishable from $z=0$ giant discs, pointing to the inefficiency of AGN feedback in halting disc growth. Using the Mapping Nearby Galaxies at Apache Point Observatory survey, we identify potential $z=0$ descendants of ADF22.1, suggesting it will evolve into an extreme (in either mass or angular momentum) early-type galaxy. Finally, we argue that cold-stream accretion, invoked to explain disc formation at $z > 1$, cannot simultaneously account for its size, dynamical properties, high specific angular momentum, and baryon-to-halo mass ratio. Instead, sustained accretion from the hot circumgalactic medium, either via spontaneous or fountain-driven condensation, offers a more physically plausible formation pathway.
\end{abstract}

\begin{keywords}
galaxies: disc -- galaxies: kinematics and dynamics -- galaxies: high-redshift
\end{keywords}



\section{Introduction}

Disc galaxies have long been regarded as a critical testbed for models of galaxy formation, evolution, and cosmology as  their structural properties (e.g., sizes, specific angular momentum) reflect the interplay between complex baryonic processes and the underlying halo assembly history \citep[e.g.,][]{Mo_1998, vandebosch_2002, Dutton_2011, Romanowsky_2012, Somerville_2018, Danovich_2015, Teklu_2015, Lagos_2018, Liang_2025b}. In the classical picture of galaxy formation, the formation of discs and their global properties are broadly explained within the framework of tidal torque theory, according to which proto-galactic density perturbations acquire angular momentum through gravitational torques exerted by the surrounding large-scale structure during the early stages of collapse before turn around \citep{Peebles_1969, White_1984, Porciani_2002}. Within this framework, the gas is assumed to initially trace the dark-matter distribution, share its specific angular momentum, and approximately conserve it as it cools, collapses, and forms stars. Despite its simplicity, this model successfully predicts that, to first order, a close correspondence should exist between the scaling relations of dark-matter haloes (e.g. size, mass, and angular momentum) and the analogous relations for disc galaxies. The most important of these is the Fall relation \citep{Fall_1983}, which links stellar or baryonic mass to specific angular momentum \citep[e.g.,][]{Fall_1983, Romanowsky_2012, Obreschkow_2014, Posti_2019, ManceraPina_2021}. 

This classical picture remains broadly valid even when complex baryonic processes are included. One important concept in this context is the notion of selective cooling (also known in the literature as biased collapse scenario), in which discs form from only a fraction of the baryons associated with their dark-matter haloes, and the collapse of gas proceeds progressively from the inner to the outer parts of the halo \citep[e.g.,][]{Dutton_2009, Dutton_2012, Kassin_2012}. As a result, higher-angular-momentum gas is accreted later and at larger radii, a process of angular momentum stratification that naturally leads to disc specific angular momenta that are comparable to, but somewhat lower than, that of the host halo. This expectation is confirmed by both simulations and observations: the specific angular momentum of disc galaxies is typically within a factor of $\sim$2 of that of their host dark-matter haloes \citep[e.g.,][]{Danovich_2015, Teklu_2015, ManceraPina_2021, Romeo_2023}.


Decades of observational studies indicate that the mass--size relation of disc galaxies evolves strongly with redshift, with high-$z$ galaxies being significantly more compact than local counterparts of similar stellar mass \citep[e.g.,][]{vanderwel_2014, Suess_2022, Martorano_2024}. In this context, the recent discovery of unexpectedly large disc galaxies at high redshift, enabled in large part by \textit{JWST}, has come as a surprise. To date, two such systems have been reported at $z>1$: the Big Wheel at $z = 3.25$ \citep{Wang_2025} and ADF22.1 at $z=3.09$
 \citep{Umehata_2025}. Both galaxies are clear outliers from the high-mass extrapolation of the mass--size relation at similar epochs \citep[e.g.,][]{Ward_2024, Allen_2025}, and their gas kinematics indicate that they are rotationally supported \citep{Rizzo_2023, Wang_2025, Umehata_2025}. Strikingly, both systems reside in overdense environments, hinting that massive, extended discs may assemble early within the highest-density regions at $z \sim 3$. Even more remarkable, these high-$z$ giant discs are so massive to resemble the most extreme disc systems known in the local Universe, the so-called super spirals \citep[hereafter referred to as giant discs; e.g.,][]{Ogle_2016, Ogle_2019}. Local giant discs are star-forming spiral galaxies with unusually high stellar masses ($\gtrsim 10^{11} M_{\odot}$), a regime in which galaxies are generally expected to be quiescent, bulge-dominated systems \citep{Zeng_2021, Jackson_2022, Pallero_2025}. 

Dynamical studies of local giant discs \citep{DiTeodoro_2021, DiTeodoro_2023} have shown that these galaxies lie on, or represent smooth extensions of, the relations followed by normal star-forming discs \citep[e.g.,][]{Lelli_2016}, including the Fall relation and the Tully--Fisher relation \citep{Tully_1977}, which connects stellar or baryonic mass to circular velocity. Similarly to normal discs, these giant discs have approximately the same baryonic specific angular momentum as their
dark-matter haloes \citep{DiTeodoro_2023}. Furthermore, rotation-curve decomposition reveals that the stellar-to-halo and baryonic-to-halo mass relations place giant discs among the most efficient galaxies in the Universe at accreting gas and converting it into stars, with stellar and baryonic mass fractions approaching the cosmological baryon fraction \citep{DiTeodoro_2023}. Specifically, they reach values of $\sim 0.3$--$0.9$ in units of the cosmological baryon fraction, $f_{\mathrm{b, cosmo}} = 0.16$, \citep{Planck}, compared to $\lesssim 0.2$--$0.3$ for early-type galaxies at similar halo masses \citep[e.g.,][]{Bilicki_2021, Posti_2021, Wang_Kai_2025}.

Observational constraints on the structural properties of high-$z$ giant discs and their 
dark-matter haloes remain scarce. No detailed dynamical study of a giant disc at high-$z$ has yet been carried out to constrain the baryonic mass distribution, the dark-matter halo mass, and the stellar and baryonic specific angular momentum, nor to place 
such systems on the dynamical scaling relations alongside their local counterparts. In the case of the Big Wheel, \citet{Wang_2025} attempted a comparison with the local stellar Tully--Fisher relation; however, this analysis relies on H$\alpha$ kinematics obtained from slit spectroscopy, which is subject to substantial uncertainties \citep{Rizzo_2024}, and no dynamical modelling was performed. For ADF22.1, \citet{Umehata_2025} analysed the kinematics using [CII] observations, but the analysis remained limited: while the stellar specific angular momentum was measured, no comparison was made with the scaling relations of local giant discs, and the dark-matter halo mass was estimated via an empirical scaling relation rather than being directly constrained through rotation-curve decomposition. 

For ADF22.1, exceptionally high-quality observations of the [CII] emission line are available \citep{Umehata_2025}. Together with complementary \textit{JWST} observations, this makes ADF22.1 a unique target for a detailed dynamical analysis. In this work, we exploit these ALMA and \textit{JWST} 
observations to characterise its dynamics, compare it to local counterparts, and use this information to understand its formation and subsequent evolution.

This paper is organised as follows. Section~\ref{sec:adf} provides an overview of ADF22.1 and the protocluster in which it resides, based on previous studies. In Section~\ref{sec:data}, we describe the data. In Section~\ref{sec:analysis}, we derive the dynamical properties of ADF22.1, including constraints on its dark-matter halo. In Section~\ref{sec:angular}, we obtain the specific angular momentum of its baryonic and dark-matter components. In Section~\ref{sec:scaling}, we discuss the location of ADF22.1 relative to local scaling relations and their expected redshift evolution. In Section~\ref{sec:descend}, we use the inferred gravitational potential to identify candidate descendant galaxies at $z=0$ and compare their properties with a mass-matched sample. In Section~\ref{sec:formation}, we discuss the most plausible formation mechanisms for ADF22.1. Finally, we summarise our main conclusions in Section~\ref{sec:conclusion}. Throughout this work, we adopt the $\Lambda$CDM cosmology from \citet{Planck}, with $H_0 = 67.6\,\mathrm{km\,s^{-1}\,Mpc^{-1}}$, 
$\Omega_{\mathrm{m}} = 0.309$, and
$\Omega_{\Lambda} = 0.688$, 
and $\Omega_{\mathrm{b}} = 0.0489$. At the redshift of our target ($z = 3.09$), this corresponds to a physical scale of $7.8\,\mathrm{kpc\,arcsec^{-1}}$.

\section{ADF22.1}
\label{sec:adf}
ADF22.1 is a spiral galaxy whose rest-frame optical and near-infrared morphology has recently been studied by \citet{Umehata_2025} using \textit{JWST} observations. In addition to being an outlier in the mass–size relation, as discussed in the Introduction, ADF22.1 hosts a remarkably extended gas disk traced by [CII] emission -- probing cool ($T \lesssim 10^3$ K) multiphase gas \citep{Wolfire_2022} -- with a diameter of $\approx 30$ kpc, and has rotation velocities, as traced by both [CII] and CO, exceeding $500$ km s$^{-1}$ \citep{Rizzo_2023, Umehata_2025}. These properties make it among the most extended and fastest-rotating galactic discs known beyond the local Universe \citep{Pensabene_2025}. Moreover, ADF22.1 is simultaneously the brightest dusty star-forming galaxy (DSFG), with a star-fomration rate (SFR) of $\sim 600$ M$_{\odot}$ yr$^{-1}$ \citep{Umehata_2025}, and the brightest X-ray active galactic nucleus (AGN) in the central region of the SSA22 protocluster complex. SSA22 was first recognized as a prominent overdensity of Lyman-break galaxies (LBGs) and Ly$\alpha$ emitters \citep{Steidel_1998, Hayashino_2004, Matsuda_2005, Yamada_2012}. Subsequent single-dish surveys at submillimetre wavelengths revealed a population of bright DSFGs across the field \citep{Tamura_2009}. Higher-resolution follow-up then showed that several of these sources resolved into multiple components: photometric studies  indicated that a significant fraction of the submillimetre sources host multiple massive stellar counterparts, suggesting they are dense galaxy groups embedded within the protocluster \citep{Kubo_2016}. Deep observations with ALMA subsequently revealed a dense $\sim$1--2~Mpc region containing at least 16 DSFGs and multiple X-ray--detected AGN, pointing to intense, coeval star formation and black-hole growth in an exceptionally rich environment \citep{Umehata_2015}. More recently, SSA22 was also shown to host extended, filamentary Ly$\alpha$ emission tracing the large-scale structure of the cosmic web \citep{Umehata_2019}.

\section{Data}
\label{sec:data}

\subsection{ ALMA}
We use observations of the 158-$\mu$m fine-structure line of ionized carbon [CII] obtained with ALMA project 2021.1.01406.S (PI: H. Umehata). We retrieve the calibrated measurement sets from the European ALMA Regional Centre \citep{Hatziminaoglou_2015}, which calibrates the raw visibilities using the standard ALMA pipeline scripts delivered with the observing packages. We then perform all subsequent processing with the Common Astronomy Software Applications package (CASA; \citealt{McMullin_2007}). We inspect the calibrated data to verify the pipeline performance and confirm that no additional flagging is required. We perform the continuum subtraction using {\sc{uvcontsub}}, modelling the continuum with a zeroth-order polynomial and subtracting it in the $uv$ plane from the measurement set containing the line spectral windows. We image the data with {\sc{tclean}} using natural weighting and produce two cubes with channel widths of 30~km~s$^{-1}$ and 100~km~s$^{-1}$, cleaning each cube down to a cleaning threshold of 1-$\sigma$, corresponding to 6 mJy and 3 mJy per spectral channel, respectively. This strategy preserves the native spectral resolution of the ALMA data while also providing a lower-resolution cube with higher signal-to-noise (S/N), allowing us to test the robustness of our kinematic measurements. The resulting cubes have a synthesized beam of $0\farcs23 \times 0\farcs27$. 

\begin{figure*} 
    \begin{center}  \includegraphics[width=\textwidth]{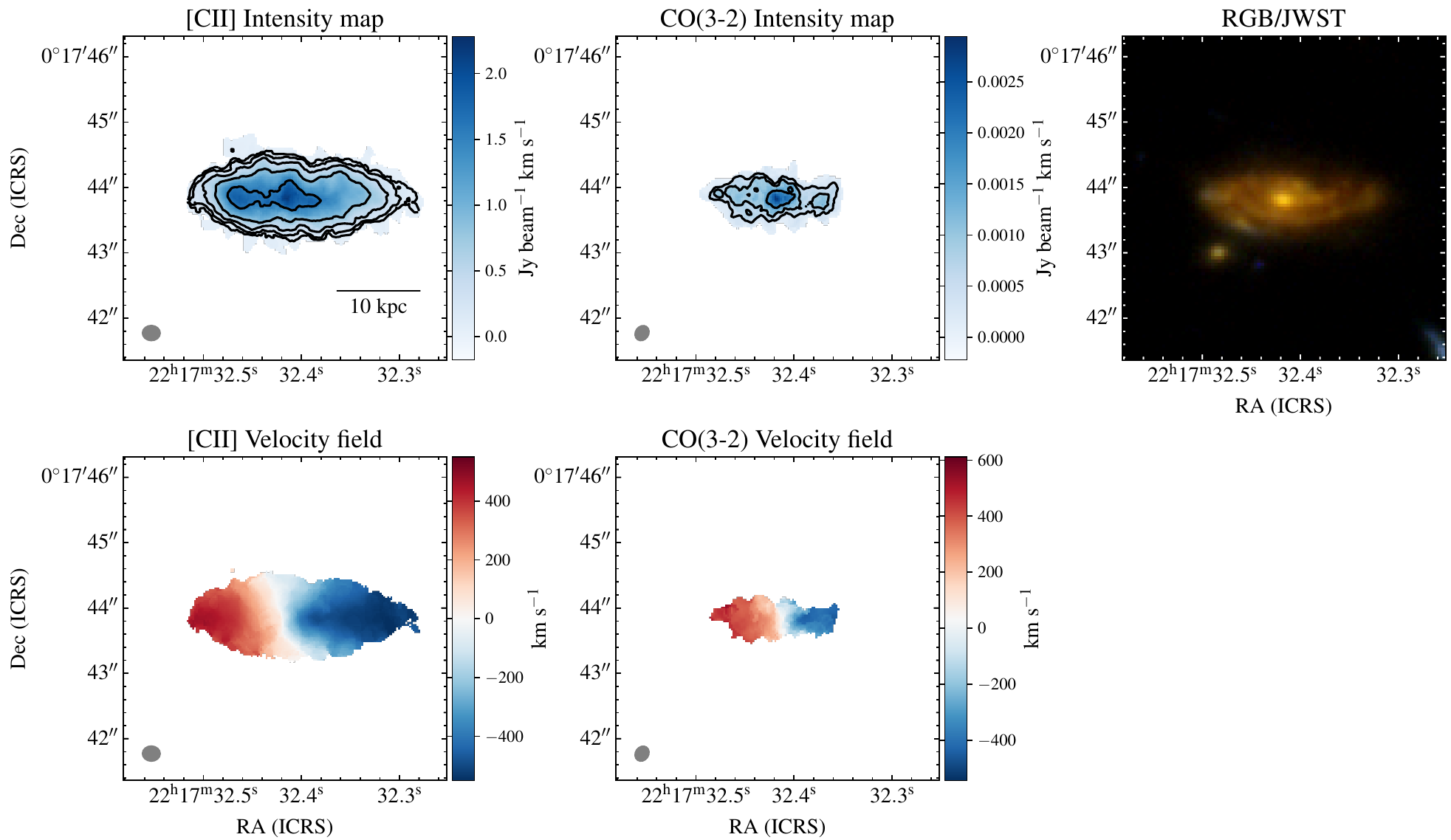}
        \caption{ALMA and \textit{JWST} imaging of ADF22.1. \textit{Upper panels}: From left to right: intensity-line maps of the 158-$\mu$m [C II] emission line, CO(3–2), and an RGB \textit{JWST} composite image (red: F444W; green: F356W; blue: F200W). The black contours show levels at $2^m \times$ the pseudo 4-$\sigma$ threshold, with $m = 0 - 6$. \textit{Bottom panels}: velocity fields derived from [CII] and CO(3-2). These images are shown for visualisation purposes only. The kinematic analysis is performed directly on the [CII] data cube, while the stellar surface-brightness analysis uses the F444W image. } 
        \label{fig:fig1}
    \end{center}
\end{figure*}

In Fig.~\ref{fig:fig1}, we show the [CII] intensity and 1st moment velocity maps. These maps are extracted using \bba\ \citep{DiTeodoro_2015}, which generates moment maps from a masked data cube. We first construct an automatic three-dimensional (3D) mask with the task {\sc{search}} in \bba, adopting SNRCUT = 3 and GROWTHCUT = 2.5. The mask is then spatially expanded by five pixels to recover low-S/N emission while minimizing contamination from the noisy background \citep{ManceraPina_2025}. Because the noise properties of the mask-integrated emission map differ from those of individual channels and are no longer spatially uniform, we compute a pseudo–4$\sigma$ contour following the method described in Appendix B of \citet{Roman_2023}. This contour is used for visualization purposes and to ensure that the selected emission is statistically significant. To build the final velocity-field map, we include only pixels above the pseudo–4$\sigma$ threshold in the intensity map.

We also include the CO(3–2) data cube from ALMA project 2018.1.01306.S (PI: H. Umehata). The imaging procedure for this dataset is described in \citet{Rizzo_2023}. The corresponding maps, derived using the same masking strategy as for the [CII] data, are shown in Fig.~\ref{fig:fig1}.

To model the surface brightness of the CO and [CII] emission (Sect.~\ref{sec:analysis}), we use the moment-0 maps obtained by collapsing the CO and [CII] data cubes along the velocity axis without applying any masking. This approach is preferred because the noise properties of the moment-0 maps are well defined and more reliably characterized than those of the line-intensity maps.

\subsection{JWST}
For ADF22.1, publicly available \textit{JWST}/NIRCam images \citep{Rieke_2005} exist in four filters (F115W, F200W, F356W and F444W) as part of a JWST Cycle-2 program (GO 3547, PI: Umehata). To trace the stellar mass distribution, we adopt the reddest available band, F444W. At $z\simeq 3$, this filter probes the rest-frame $\sim$1.1~$\mu$m emission, which reduces the impact of dust attenuation while exploiting the exquisite angular resolution of 0.14'' of \textit{JWST}/NIRCam. Emission at these wavelengths is expected to more closely trace the stellar mass surface brightness than rest-frame optical light \citep{Suess_2022, Martorano_2024}. We retrieve the reduced images from the DAWN JWST Archive (DJA), which provides processed public \textit{JWST} imaging products \citep{Valentino_2023, Heintz_2025}. In Fig.~\ref{fig:fig1}, for visualisation purposes only, we show an RGB composite image from the F444W, F356W and F200W filters. 


\section{Analysis}
\label{sec:analysis}
In this section, we model the surface brightness of the stellar and gas components of ADF22.1 (Section \ref{sec:sb}), as well as its kinematics (Section \ref{sec:kinematic}) and dynamics (Section \ref{sec:dynamic}). We then present the best-fitting models obtained by applying these methods to the observations described in Section \ref{sec:data}. The resulting best-fitting parameters are used in Sections \ref{sec:scaling} and \ref{sec:descend} to examine where ADF22.1 lies relative to local and $z = 3$ scaling relations.

\subsection{Surface brightness modelling}
\label{sec:sb}
To fit the surface brightness of the stellar and gas components using \textit{JWST} and ALMA observations, we use the ``Source Characterization using a Composable Analysis'' library \citep[{\sc{socca}}; ][]{DiMascolo_2026}. The code {\sc{socca}} is designed for fitting 2D maps, allowing users to specify the number of components and type of profiles for each component (e.g., Sérsic, exponential) and convolving the profiles with a Gaussian or a user-provided PSF. The code leverages the JAX framework for just-in-time compilation and uses the nested sampling algorithm {\sc{nautilus}}\footnote{\url{https://github.com/GNOME/nautilus-python}} \citep{nautilus} to explore the posterior distribution efficiently. As detailed below, we fit the data for our target using Sérsic \citep{Sersic_1963} and polyexponential profiles \citep{Bacchini_2019}, as defined by 

\begin{equation} \label{eq:sersic}
    I(R) = I_\mathrm{e} \exp \left\{ -b_n \left[ \left(\frac{R}{R_\mathrm{e}}\right)^{1/n} - 1 \right] \right\},
\end{equation}
\begin{equation} \label{eq:poly}
    I(R) = \exp \left( -\frac{R}{R_\mathrm{d}} \right) \left[ I_0 + I_1 \left(\frac{R}{R_\mathrm{d}} \right) + I_2 \left(\frac{R}{R_\mathrm{d}} \right)^2  \right],
\end{equation}

In equation~(\ref{eq:sersic}), $I_\mathrm{e}$ is the surface brightness at $R=R_\mathrm{e}$, where $R_\mathrm{e}$ is the effective radius, containing half the total luminosity, $n$
is the Sérsic index, controlling the shape of the profile and $b_n$ is a function of the Sérsic index \citep{Ciotti_1999}. In equation~(\ref{eq:poly}), $I_0$ is the central surface brightness at $R = 0$, $R_\mathrm{d}$ is the scale radius of the disc, and $I_\mathrm{i}$ are the coefficients describing the polynomial function. The polyexponential profile is commonly adopted to model gas surface-brightness distributions \citep{Bacchini_2019, ManceraPina_2021b}; we follow this approach and use it for fitting the CO and [CII] maps. In addition to the parameters defining the surface brightness profiles, {\sc{socca}} also enables fitting the geometrical parameters of the galaxy, including its centre ($x_0$, $y_0$), position angle ($PA$), and axis ratio ($b/a$).

\subsubsection{Deriving the stellar surface brightness using JWST data}
To model the surface brightness of ADF22.1, we first create a cutout of the \textit{JWST}/F444W filters centred on the target (RA: 334.38507 deg, Dec: 0.29551 deg) and with a size of 5'' $\times$ 5''. We generate the F444W PSF using the package STPSF \citep{Perrin_2014}. To model the cutout containing our target, we adopt a three-component model: two components represent ADF22.1, while a third accounts for the compact source located southeast of ADF22.1 (see Fig.~\ref{fig:f444}). The additional source to the southwest is excluded from the fit by masking it out. We adopt this strategy because the southeastern source lies in close proximity to ADF22.1, so masking it would require arbitrary assumptions about how the blended flux should be partitioned between the two objects. In contrast, the southwestern source is sufficiently separated that it can be cleanly removed with a mask without affecting the fit.

We initially modelled the surface brightness of our target using two Sérsic components: an inner component with a free Sérsic index and an outer, extended component with the Sérsic index fixed at 1 (i.e., an exponential disc). To differentiate between the two components, we parameterized the effective radius of the inner component as a fraction 
$k$ of the effective radius of the outer component, adopting a flat prior on 
$k$ in the range [0, 1]. However, during fitting, the Sérsic index of the inner component converged to the lower bound of the prior range ($n=0.2$), and the modelled disc appeared more extended than the observed disc. This suggests a possible shortcoming in the adopted model. To address this, we refit the data allowing both Sérsic indices to vary freely. In this case, the Sérsic index of the outer component reached the edge of its prior range, again indicating limitations in the model. As a compromise, we adopted a model with a free Sérsic index for the inner component, assumed to be associated with a bulge, and a fixed Sérsic index of 0.5 for the outer component, assumed to be associate with the disc. The parameters defining the best-fitting model are summarised in Table \ref{tab:tab1}. The resulting luminosity ratio between the bulge and total (bulge + disc) is 0.13.

In Fig.~\ref{fig:f444}, we present the two-dimensional data, the corresponding best-fitting model, and the residual map, along with the one-dimensional surface brightness profiles obtained by azimuthally averaging the flux within elliptical annuli. The annuli share the same geometry (centre, position angle, and axis ratio) as the best-fitting disc model and have a width of 0.1'', approximately matching the full width at half maximum (FWHM) of the PSF. This 1D profile is shown for visualisation purposes, as the fit is performed in two dimensions. The residual map reveals some systematic features associated with the presence of spiral arms, which are not accounted for in our model. Although locally significant, the positive and negative residuals on average cancel each out in the 1D profile, which is well described by our axisymmetric model. 

\begin{figure*} \label{fig:f444}
    \begin{center}  \includegraphics[width=\textwidth]{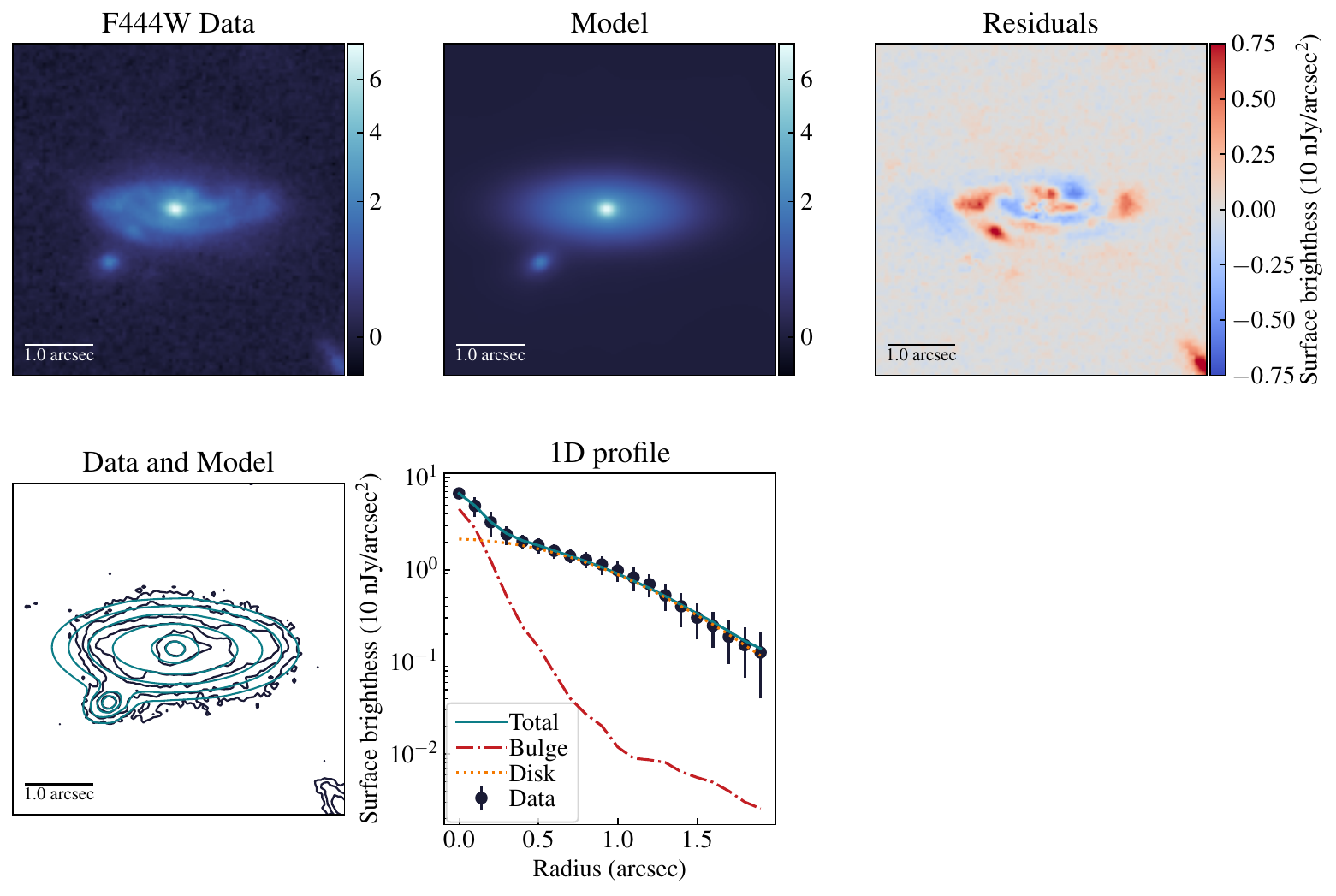}
        \caption{Stellar surface brightness map and modelling from the F444W filter. \textit{Upper panels:} from left to right: data, best-fitting model, and residual map. The residuals are shown with a different colour scale to highlight their amplitude. \textit{Bottom panels:} isophotal contours of the data (black) and model (teal) at levels of $3\times2^m$ rms, with $m = 0$–5 (left), and the 1D radial surface-brightness profile (right). The profile shows the data (circles) and the best-fitting models for the bulge (dot–dashed), the disc (dotted), and the total (disc+bulge; solid). The 1D profile is displayed for visualisation only, as the fit is performed in two dimensions.}  
        \label{fig:f444}
    \end{center}
\end{figure*}

\subsubsection{Deriving the gas surface brightness using ALMA data}
We modelled the moment-0 maps derived from CO(3-2) and [CII] data using a polyexponential profile. Figs \ref{fig:cii} and \ref{fig:co} display the data and model maps, along with the corresponding 1D surface brightness profiles extracted from the data and the model in elliptical rings with width comparable to the beam size. The best-fitting parameters describing the CO(3-2) and [CII] model profiles are summarised in Table \ref{tab:tab1}. We note that the scale length derived from the polyexponential fit should not be interpreted as a direct measure of the spatial extent of the galaxy. The polyexponential form is adopted to reproduce the observed surface brightness profile as accurately as possible, rather than to provide a size estimate directly comparable to the exponential scale lengths typically used for structural size measurements.

\begin{figure*}\label{fig:cii}
    \begin{center}  \includegraphics[width=\textwidth]{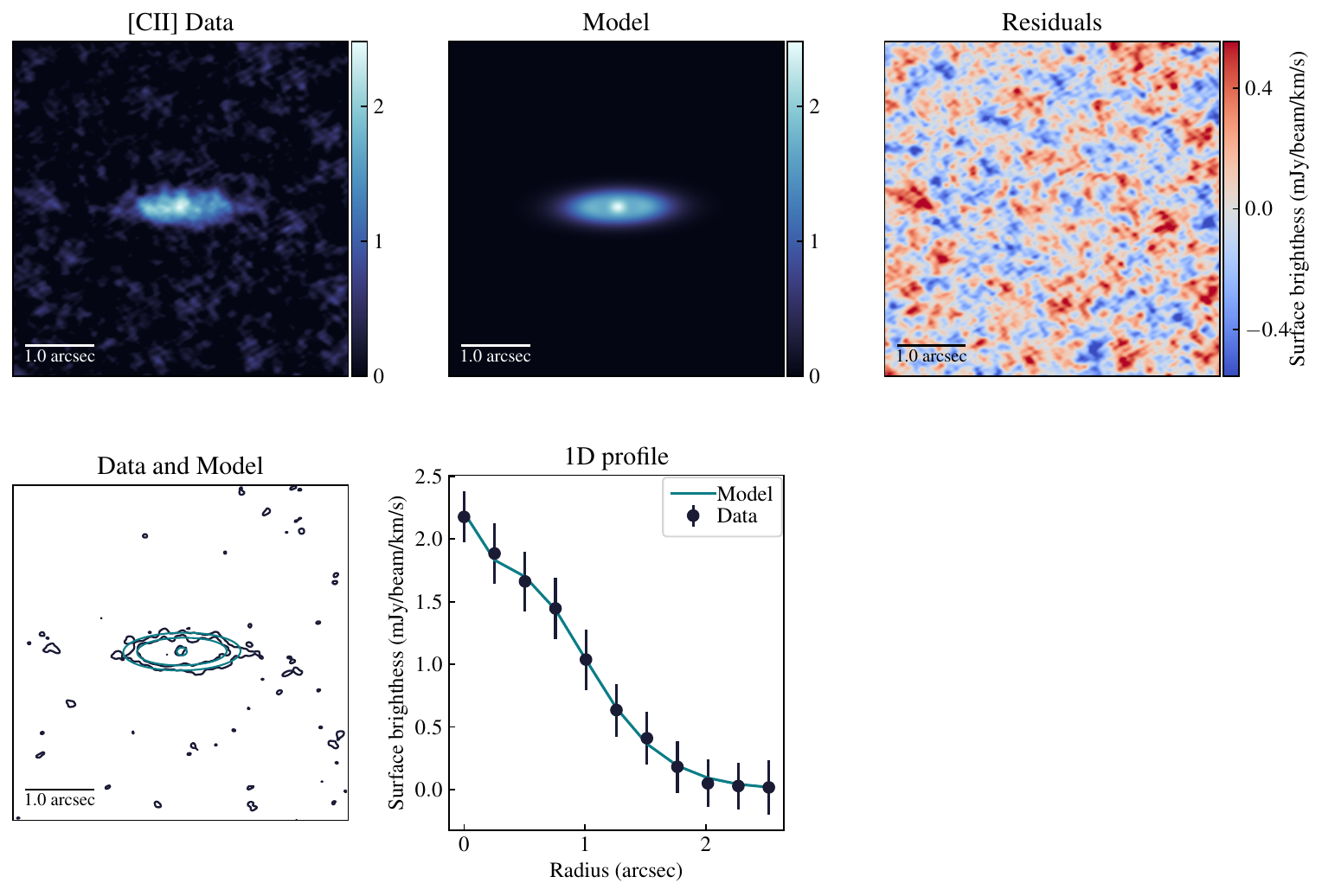}
        \caption{Same as Fig.~\ref{fig:f444}, but for [CII] data.}  
        \label{fig:cii}
    \end{center}
\end{figure*}

\begin{figure*} \label{fig:co}
    \begin{center}  \includegraphics[width=\textwidth]{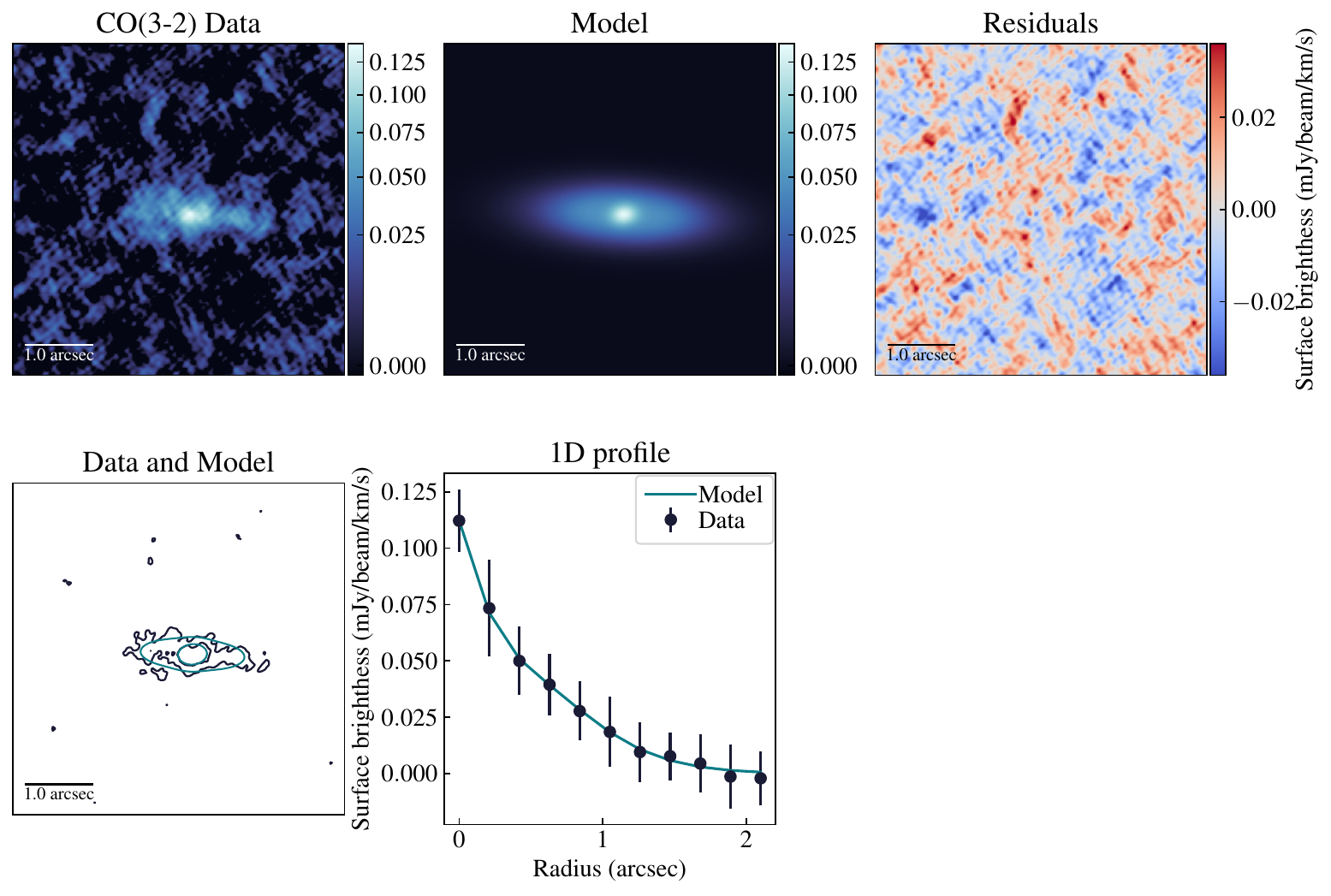}
        \caption{Same as Fig.~\ref{fig:f444}, but for CO(3-2)}  
        \label{fig:co}
    \end{center}
\end{figure*}

\begin{table*}
\begin{center}
\caption{Best-fitting parameters for the \textit{JWST}/F444W (upper table) and the ALMA [CII] and CO moment-0th maps. For the \textit{JWST} data the units $f$ are 10 nJy/arcsec$^2$, while for ALMA data they are mJy/km/s/arsec$^2$. $^*$The global effective radius derived from the combined two S\'ersic components is $6.2 \pm 1$ kpc.
}
\label{tab:tab1}
\begin{tabular}{c|ccccccccc}
\hline\hline \noalign{\smallskip}
Data & Type &  \multicolumn{6}{c}{Parameters} &\\
\noalign{\smallskip}
\hline
\noalign{\medskip}
\textbf{\textit{JWST}/F444W} & Sérsic$^*$ & $I_{\mathrm{e}}$ ($f$) & $R_{\mathrm{e}}$ (arcsec) & $n$ & $PA$ (deg) & $b/a$ \\
 &  & $0.79^{+0.01}_{-0.01}$ & 0.095$^{+0.001}_{-0.001}$ & 0.69$^{+0.06}_{-0.06}$ & 86$^{+1}_{-1}$ & 0.66$^{+0.02}_{-0.01}$\\
 \noalign{\smallskip}
 &  & 
 $0.099^{+0.001}_{-0.001}$ & 0.88$^{+0.01}_{-0.01}$ & 0.5 & 88$^{+1}_{-1}$ & 0.37$^{+0.01}_{-0.01}$ \\
\noalign{\smallskip}
\hline
\noalign{\medskip}
 & Polyexp & $I_{\mathrm{0}}(f)$ & $R_{\mathrm{d}}$ (arcsec) & $I_1$ ($f$) &  $I_2$ ($f$) & $PA$ (deg) & $b/a$\\
\textbf{ALMA/[CII]} & & $129^{+4}_{-4}$ & 0.240$^{+0.001}_{-0.001}$ & $-170^{+8}_{-6}$  & 84$^{+2}_{-2}$  & 90.9$^{+0.1}_{-0.1}$ & 0.295$^{+0.002}_{-0.002}$\\
\noalign{\smallskip}
\textbf{ALMA/CO} &  & 7.2$^{+0.3}_{-0.3}$  & 0.203$^{+0.003}_{-0.003}$  & $-7.0^{+0.4}_{-0.4}$ & 2.6$^{+0.1}_{-0.1}$  & 86.5$^{+0.3}_{-0.3}$ & 0.300$^{+0.005}_{-0.005}$
\\
\noalign{\smallskip}
\hline
\end{tabular}
\end{center}
\end{table*}

\subsection{Kinematic modelling}
\label{sec:kinematic}
To model the kinematics of ADF22.1, we fit the high- and low-spectral-resolution data cubes of [CII] using \bba\ \citep{DiTeodoro_2015}. The kinematic model implemented by \bba\ is based on a tilted-ring approach, where the galaxy is divided into concentric rings. Each ring is characterized by the kinematic parameters (i.e., rotation velocity $V_{\mathrm{rot}}$, velocity dispersion $\sigma$, and systemic velocity), and geometrical properties, including the centre, inclination $i$, and kinematic position angle $PA_\mathrm{kin}$. \bba\ employs a forward modelling technique, which involves constructing the model in 3D with the same specifications as the observational data cubes (e.g., pixel scale and channel width). The model is then convolved with the observational beam to match the resolution before residuals are computed.

To model the high- and low-spectral-resolution data, we first fix geometrical parameters (centre and inclination) to the ones obtained in the surface brightness fitting. Specifically, the centre is fixed to the best-fitting parameters obtained from fitting the [CII] moment-0 map, and the inclination is set based on the axis ratio derived from the [CII] surface brightness fit, assuming a thin disc geometry (i.e., $\cos i = b/a$). The resulting inclination is 72$^\circ$, which differs by 4$^\circ$ from the value of 68$^\circ$ obtained using the axis ratio of the \textit{JWST}/F444W data, under the same assumption of a thin disc geometry. To account for potential effects of the inclination on the derived kinematic parameters, we constructed two sets of models for each data cube: one with the inclination fixed to 72$^\circ$ and another with the inclination fixed to 68$^\circ$. For the cube fitting, we adopt the mask generated with the {\sc search} task and spatially expanded by five pixels, as described in Section~\ref{sec:data}. For the rotating disc model built by \bba\, we assume the {\sc azim} normalization and perform the analysis in two steps. In the first step, the rotation velocity, velocity dispersion, position angle, and systemic velocity are allowed to vary freely. In the second step, the systemic velocity is fixed, and the rotation velocity, velocity dispersion, and position angle are refitted to reduce the number of free parameters. The \bba\ model includes a total of 9 rings for the low-spectral-resolution cubes and 8 rings for the high-spectral-resolution cubes, each with a width of 0.23", equal to the minor axis of the beam. This ensures that the parameters for each ring are nearly independent. The different number of rings adopted in the low- and high–spectral-resolution cases reflects the higher S/N of the low–spectral-resolution data, which allows an additional ring to be reliably fitted. 

\begin{figure*} 
    \begin{center}  \includegraphics[width=\textwidth]{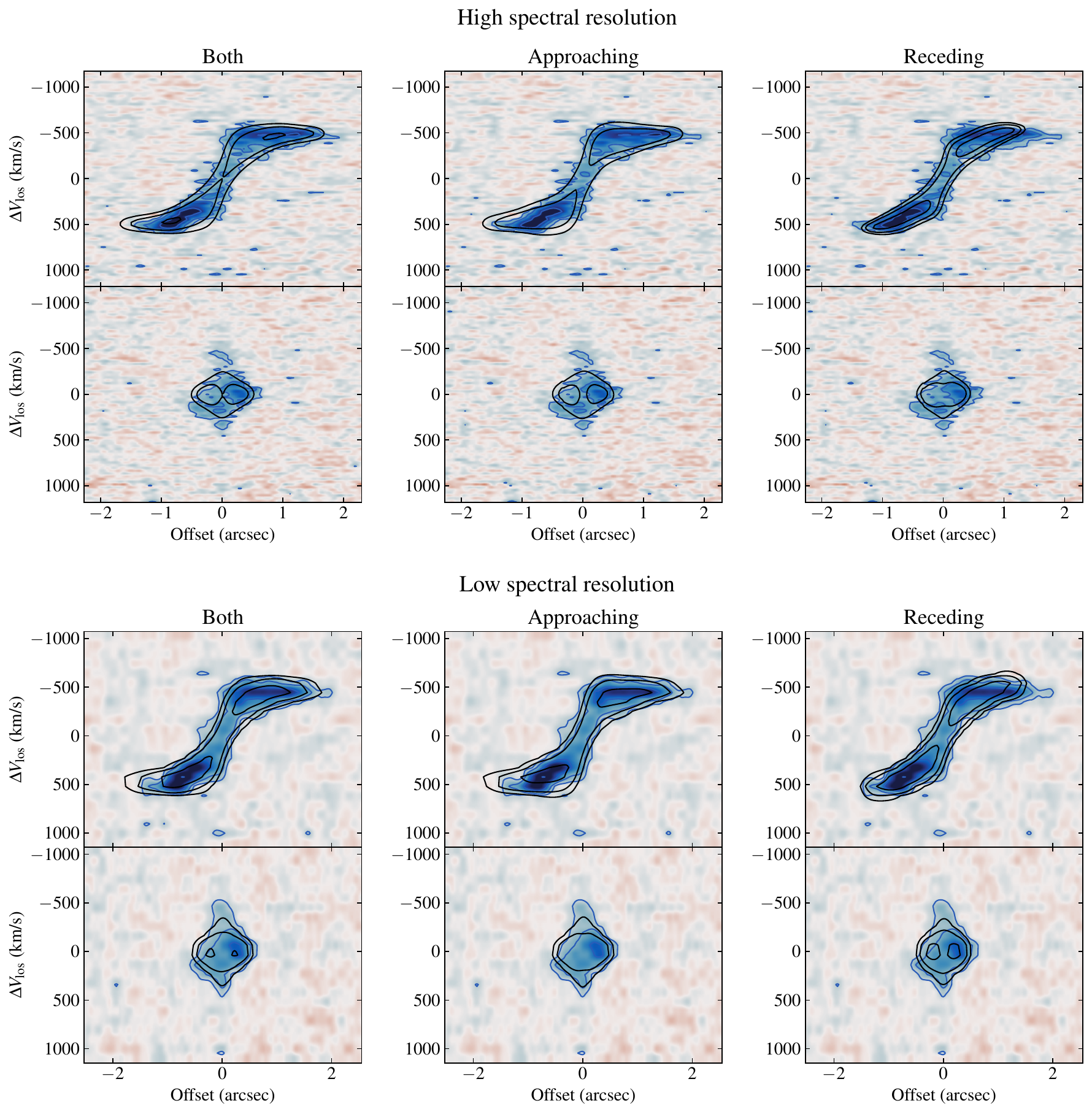}
        \caption{Position–velocity diagrams extracted along the major and minor axes, showing the data (black contours) and the model (blue contours).  The contours follow levels of $3\times2^m$ rms, with $m = 0$–4. The top row presents the high spectral-resolution dataset, and the bottom row the low spectral-resolution dataset. From left to right: model obtained by fitting the full galaxy, and models obtained by fitting separately the approaching and receding sides.}  
        \label{fig:pv}
    \end{center}
\end{figure*}

The models obtained with fixed inclinations of 72$^\circ$ and 68$^\circ$ yield consistent results. Consequently, throughout the remainder of this paper, we present only the results obtained with the inclination fixed at 72$^\circ$. In Fig. \ref{fig:pv} (first column), we present the position-velocity diagrams of the data (blue contours) and the model (black contours) along the major and minor axes for both the high- and low-spectral-resolution data. Overall, the \bba\ models successfully reproduce the bulk of the ADF22 emission. However, the approaching side of the galaxy appears more extended and is reproduced more faithfully than the receding side. Such differences between the approaching and receding halves, often referred to as lopsidedness, are also common in spiral galaxies in the local Universe \citep{Baldwin_1980, Richter_1994, Haynes_1998}. The physical origin of lopsidedness remains debated \citep[e.g.,][]{Bournaud_2005, Sancisi_2008, Jog_2009, Lokas_2022, Dolfi_2023}. \citet{Umehata_2025} attributed the observed asymmetry to the presence of a bar within the inner $\sim$3 kpc. However, while lopsidedness can in some cases be associated with bars, we find no clear evidence for bar-driven non-circular motions in the [CII] kinematics, nor for a bar-like feature in the stellar surface-brightness distribution. Although asymmetries in the minor-axis PV diagram suggest the presence of non-circular motions, the current data and the inclination of the galaxy with respect to the line of sight do not allow us to robustly determine whether these can be attributed to a bar.

\begin{figure*} 
    \begin{center}  \includegraphics[width=\textwidth]{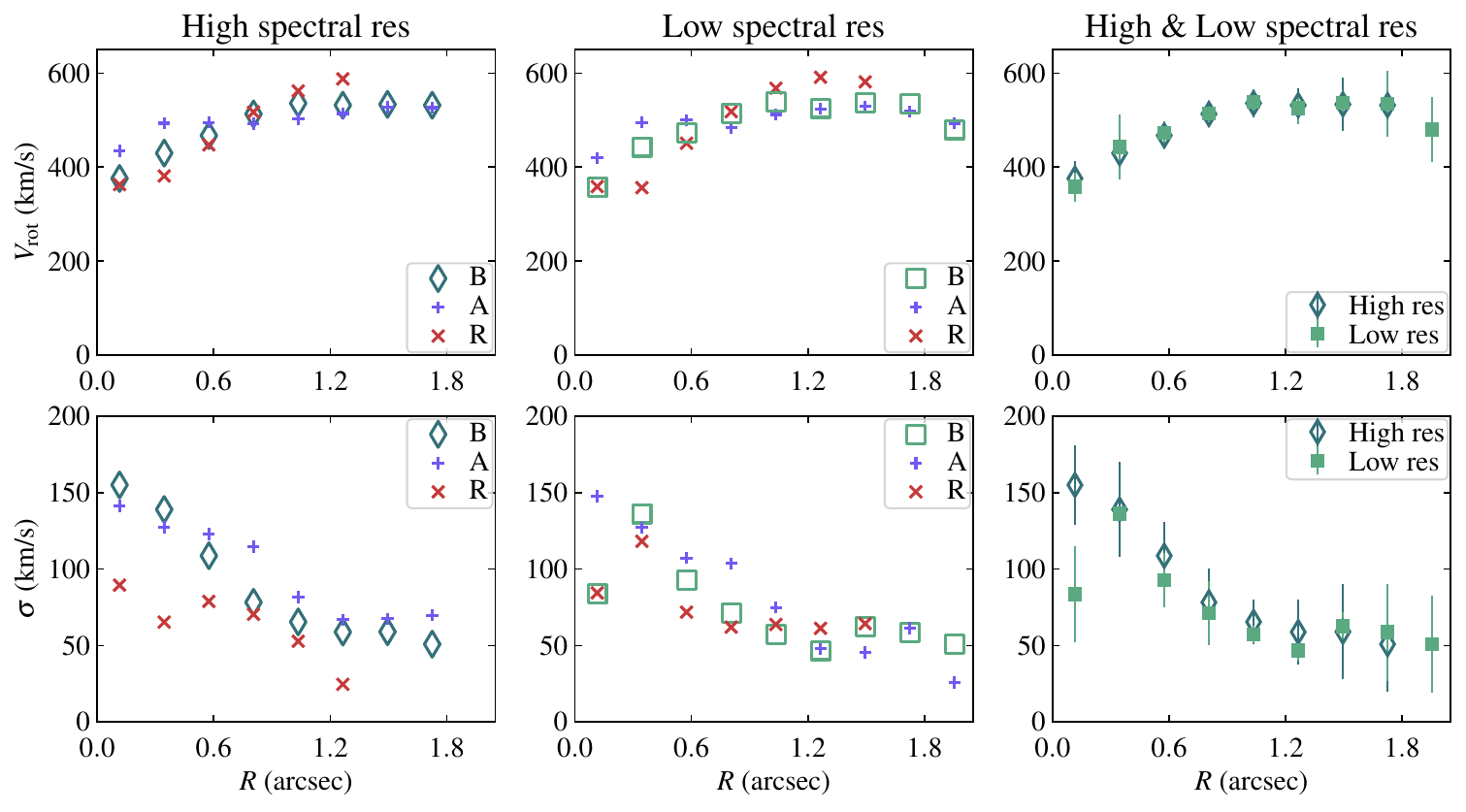}
        \caption{Best-fitting rotation curves (top panels) and velocity-dispersion profiles (bottom panels) derived from fits to the high–spectral-resolution (diamonds) and low–spectral-resolution (squares) data cubes, modelling the full galaxy, as well as fitting the approaching (pluses) and receding (crosses) sides separately.}  
        \label{fig:comparison}
    \end{center}
\end{figure*}

To account for the kinematics lopsidedness, we separately model the approaching and receding sides of the galaxy (see the second and third columns in Fig. \ref{fig:pv}). In Fig. \ref{fig:comparison}, we show the profiles of the rotation velocity and velocity dispersion obtained when fitting the full galaxy and when considering only the approaching or receding sides, for both the low- and high-angular-resolution cases. The median difference in the rotation velocity between the approaching and receding sides is of the order of 3\% and 7\% for the high- and low-spectral-resolution data sets, respectively. The largest differences, reaching up to 26\% and 31\%, occur in the inner regions (e.g., second point of the profiles). This is primarily due to the approaching side having a steeper increase in the rotation velocity, while the receding side shows an unusual initially flat central rotation velocity profile, followed by a steep increase starting from the third resolution element. The differences in the velocity dispersion between the approaching and receding side fits are more pronounced, with median values of 19\% and 44\% for the low- and high-spectral-resolution data, respectively. This is because the \bba\ model attempts to account for asymmetries, which are particularly prominent in the inner regions. For instance, the minor axis is notably asymmetric with respect to the centre and systemic velocity, displaying an elongated emission feature near the inner regions and at negative velocities that the model does not fully reproduce. Additionally, the velocity dispersion is more sensitive to noise \citep{Rizzo_2022}, which is higher in the high-spectral-resolution data, further contributing to the observed discrepancies. 

Unless otherwise stated, we adopt the model obtained by fitting the full galaxy as our fiducial model. To account for differences in measurements between the approaching and receding sides, we follow the method outlined by \citet{Swaters_2009}, where the uncertainties on the fiducial $V_{\mathrm{rot}}(R)$ and $\sigma(R)$ are defined as half the difference between the parameters of the two sides, i.e., $(V_{\mathrm{rot, app}}(R) - V_{\mathrm{rot, rec}}(R))/2$. However, unlike \citet{Swaters_2009}, who interpreted this quantity as a 2-$\sigma$ uncertainty, we adopt a more conservative approach and treat it as a 1-$\sigma$ uncertainty. For the last two points, where only the approaching side fit is available due to its more elongated structure, we adopt a conservative approach by assigning the maximum errors observed in the preceding points. When considering the velocities for the approaching and receding sides, we use the same uncertainties as those obtained for the full galaxy model.

The comparison between the low- and high-spectral-resolution fits (third column in Fig.~\ref{fig:comparison}) shows that the rotation velocities and velocity dispersions are generally in full agreement. The only exception is the first point in the velocity dispersion, which differs by 1.5-$\sigma$ (see Section \ref{sec:vc} for further details). For the remainder of the paper, we will base our analysis on the fits obtained from the high-spectral-resolution data.

The rotation curve of ADF22.1 reaches velocities of $\sim 530$ km s$^{-1}$ in the outskirts, confirming previous measurements based on the more compact CO(3-2) emission \citep{Rizzo_2023} and the recent [CII] kinematic analysis by \citet{Umehata_2025}. Such high rotation speeds are rare among spiral galaxies at $z = 0$ \citep{Lelli_2016}, but have been increasingly reported in recent years for massive, dusty galaxies at high redshift \citep[e.g.,][]{Fraternali_2021, Rizzo_2021, Roman_2023, Rizzo_2023, Pensabene_2025, Kaasinen_2026}.

\subsection{Dynamic modelling}
\label{sec:dynamic}

\subsubsection{Deriving the circular velocity} \label{sec:vc}
The circular speed of a galaxy ($V_{\mathrm{c}}$) is a direct tracer of the total gravitational potential $\phi$. Once the gas rotation velocity ($V_{\mathrm{rot}}$) and velocity dispersion ($\sigma$) are available, deriving the circular velocity $V_{\mathrm{c}}$ becomes a relatively straightforward process. For a thin gas disc, the following equation holds \citep{Cimatti_2019}:
\begin{equation} \label{eq:vc} V_{\mathrm{c}}^2 = R \frac{\partial \phi}{ \partial R} = V_{\mathrm{rot}}^2 - R \sigma^2 \frac{\partial \ln (\sigma^2 \Sigma)}{\partial R} = V_{\mathrm{rot}}^2 + V_{\mathrm{A}}^2, \end{equation}

where $\Sigma$ represents the gas mass surface density distribution. Since $\Sigma$ enters the equation as a logarithmic derivative, the [CII] surface brightness profile derived in Section \ref{sec:sb} can be directly used. As a result, this calculation is entirely independent on the $\alpha_{\mathrm{[CII]}}$ conversion factor (see Section \ref{sec:massdec}). The term $V_{\mathrm{A}}$, known as the asymmetric drift correction, accounts for the pressure support provided by the random motions of gas. This correction is particularly significant in regions where the velocity dispersion is comparable to or exceeds the rotation velocity.

We calculate the circular velocity under three scenarios:

\begin{itemize}

\item We apply Equation (\ref{eq:vc}) using the rotation velocity and velocity dispersion derived from fitting the full galaxy. We refer to the resulting circular-velocity profile as the fiducial profile.

\item Due to the lopsidedness in ADF22.1, there are differences in $V_{\mathrm{rot}}$ and $\sigma$ between the approaching and receding sides. We thus derive the circular velocity using the fit for the receding and approaching sides separately. 

\item For reference, we also consider the scenario where $V_{\mathrm{c}} \approx V_{\mathrm{rot}}$. This approximation is valid everywhere except at $R \lesssim 5$ kpc, where, however it was already noted that the velocity dispersion may have been overestimated (Section \ref{sec:kinematic}).

\end{itemize}

\begin{figure}
    \begin{center}  \includegraphics[width=\columnwidth]{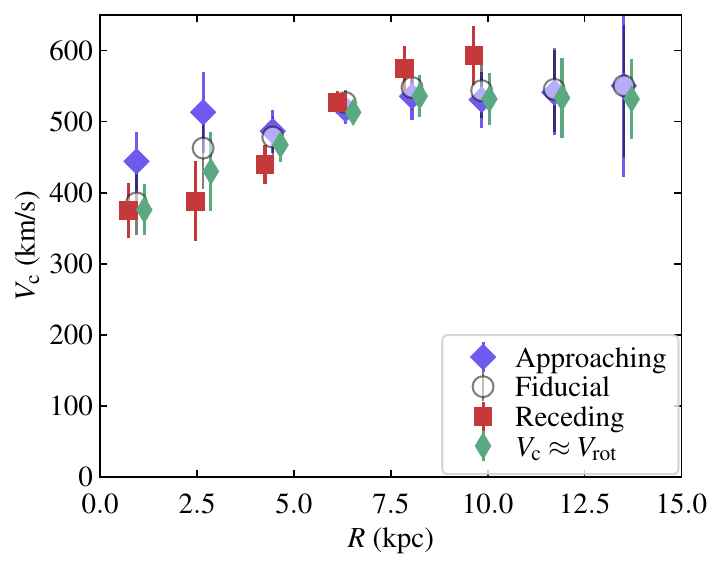}
        \caption{Comparison of circular-velocity profiles derived from the fiducial rotation curve and velocity dispersion profile (circles), the approaching (diamonds) and receding (squares) sides separately, and the case where the observed rotation velocity is assumed to be a good approximation of the circular velocity (thin diamonds).}  
        \label{fig:vc}
    \end{center}
\end{figure}

In Fig.~\ref{fig:vc}, we present the circular-velocity profiles derived under the four assumptions described above. The resulting circular velocities are consistent with one another within the uncertainties. The largest discrepancies occur in the first two radial points. We account for this when comparing the fiducial circular velocity profile with the model from the rotation-curve decomposition as shown below.

\subsubsection{Rotation-curve decomposition} \label{sec:massdec}

In this section, we present a mass decomposition model for the circular velocity of ADF22.1, following the assumptions employed in recent ALMA and \textit{JWST} studies of galaxy rotation curves at $z > 1$ \citep{Roman_2026}. We write the model circular velocity, $V_{\mathrm{tot}}(R)$, as the sum in quadrature of contributions from both baryonic and dark matter components:

\begin{equation}
    V^2_{\mathrm{tot}}(R) = V^2_{\mathrm{b}}(R) + V^2_{\mathrm{DM}}(R)
\end{equation}

The baryonic contribution, \(V_{\mathrm{b}}(R)\), includes both gas and stellar components. We model the gas as the sum of dense molecular and diffuse components:
    \begin{itemize}
        \item The dense molecular gas (hereafter Gas$_{\mathrm{mol}}$) is assumed to follow the same spatial distribution as the CO(3--2) emission. The total mass for the molecular gas is assumed to be given by the luminosity of CO(1--0) of 8 $\times 10^{10}$ K km s$^{-1}$ pc$^2$ \citep{Umehata_2025} normalized by a CO-to-H$_2$ conversion factor (\(\alpha_{\mathrm{CO}}\)), which is treated as a free parameter.
        \item The diffuse gas (hereafter Gas$_{\mathrm{dif}}$) is assumed to trace the spatial distribution of the [CII] emission, with a normalization set by scaling the [CII] luminosity of 1.2 $ \times 10^{10} L_{\odot}$ via a free parameter \(\alpha_{\mathrm{CII}}\).
    \end{itemize}
    The surface density profiles of both gas components are defined using the parameters of the polyexponential profiles found in Section \ref{sec:sb}. 

    The stellar component is modelled based on the best-fitting parameters (Sérsic index \(n\), effective radius $R_{\mathrm{e}}$) derived from the modelling of the F444W cutout. Specifically, we assume a spherical stellar component whose two-dimensional projection reproduces the Sérsic profile describing the inner component of the F444W model, and a thin disc geometry for the outer component.

    Although the original bulge-to-total (B/T) light ratio derived from F444W is 0.13, we adopt a B/T of 0.2 based on the spatially resolved SED fitting by \citet{Umehata_2025}. This choice implies a bulge mass-to-light ratio 1.7 times higher than that of the disc, consistent with the expected differences in dust and stellar population distributions between the central and outer regions \citep{Roman_2026, ManceraPina_2025}. We model the total stellar mass as $M_\star = \alpha_\star \, M_{\star, \mathrm{SED}}$, where
     $\log(M_{\star, \mathrm{SED}}/M\odot) = 11.4$ is the value derived from SED fitting by \citet{Umehata_2025}, and $\alpha_\star$ is a normalization that is left free to accommodate small variations and potential systematic uncertainties with respect to the SED value.

The dark matter contribution, \(V_{\mathrm{DM}}(R)\), is modelled using a Navarro-Frenk-White (NFW) profile \citep{Navaro_97} characterized by two parameters: the halo mass \(M_{200}\) and the concentration \(c_{200}\). We note that $M_{200}$ and $c_{200}$ do not correspond to the true virial quantities, since at the redshift of our target the virial overdensity is $174$ rather than $200$ \citep{Bryan_1998}. Nevertheless, we adopt the $200$ definition to maintain consistency with the $z = 0$ literature \citep[e.g.,][]{Posti_2019, DiTeodoro_2021, ManceraPina_2025}. At redshifts \(z \gtrsim 3\), the mass--concentration relation is nearly flat, and haloes with \(M \gtrsim 10^{10} \, M_{\odot}\) typically have concentrations of 3--4 \citep[e.g.,][]{Dutton_2014, Child_2018}, with a typical scatter of 0.13 dex. Accordingly, we fix \(c_{200} = 3.5\) and fit only for \(M_{200}\).

Overall, the free normalization parameters in the model are: \(\alpha_{\mathrm{CII}}, \alpha_{\mathrm{CO}}, \alpha_{\star}\), and \(M_{200}\). The total model circular velocity is then given by:

\begin{align}  
    V_{\mathrm{tot}}(R) = \bigg[ \, 
    & \alpha_{\mathrm{[CII]}} V^2_{\mathrm{CII}}(R \mid p_{\mathrm{i, [CII]}}) +
        \alpha_{\mathrm{CO}} V^2_{\mathrm{CO}}(R \mid p_{\mathrm{i, CO}}) \nonumber \\ 
        & + \alpha_{\star} V^2_{\star}(R \mid p_{\mathrm{i}, \star}) +
        V^2_{\mathrm{DM}}(R, M_{200} \mid p_{\mathrm{i, DM}}) \, \bigg]^{1/2}
\end{align}

Here, \(p_{\mathrm{i}}\) define the parameters that are kept fixed:
\begin{itemize}
    \item \(R_\mathrm{d}, I_0, I_1, I_2\) defining the polyexponential profiles used for Gas$_{\mathrm{mol}}$ and Gas$_{\mathrm{dif}}$, derived from CO and [CII] observations.
    \item \(n, R_\mathrm{e}, B/T\) defining the Sérsic profiles describing the stellar bulge and disc.
    \item \(c_{200}\) for the dark-matter halo.
\end{itemize}

The model velocities for each component are computed using the {\sc{galpynamics}} code\footnote{https://gitlab.com/iogiul/galpynamics/}, and the best-fitting parameters are inferred using the Bayesian sampler {\sc{dynesty}} \citep{Speagle_2020, Koposov_2025}. The priors for all free parameters are summarised in Table~\ref{tab:tab_prior}. For the rescaling factor $\alpha_{\star}$ and $\log (M_{200})$ we adopt uniform flat priors. In the case of $\log (M_{200})$, the lower bound is set by the cosmological baryon fraction of 0.16 ensuring that the halo mass is consistent with the minimum required to host the observed baryonic content. For $\alpha_{\mathrm{CO}}$ and $\alpha_{\mathrm{CII}}$, we adopt truncated normal priors centered on $\alpha_{\mathrm{CO}} = 2.5\,M_\odot\,(\mathrm{K\,km\,s^{-1}\,pc^2})^{-1}$ and $\alpha_{\mathrm{CII}} = 20\,M_\odot L_\odot^{-1}$, with standard deviations of 1.5 and 10, respectively. The adopted central values are representative of those commonly assumed for high-$z$ galaxies, while the relatively broad dispersions reflect the substantial systematic uncertainty and the lack of consensus on their values \citep[e.g.,][]{Daddi_2010, Zanella_2018, Tacconi_2020, Kirkpatrick_2019, Madden_2020, Rizzo_2021}. The distributions are truncated at zero to enforce physically meaningful positive values.

Table~\ref{tab:dyn_bestfit} reports the best-fitting values and derived quantities, such as the gas mass  (i.e., $M_{\mathrm{gas}} = M_{\mathrm{gas, mol}} + M_{\mathrm{gas, dif}}$), the total baryonic mass (i.e., $M_{\mathrm{b}} = M_{\star} + M_{\mathrm{gas}}$), stellar-to-dark matter mass ratio, and baryonic-to-dark matter mass ratio. The posterior distributions are shown in Fig.~\ref{fig:corner} in the Appendix. Fig.~\ref{fig:decomposition_fiducial} shows the fiducial circular velocity curve and the best-fitting model. In Fig.~\ref{fig:decomposition_comp} in the Appendix, we show the comparison between the different circular velocity profiles derived in Section \ref{sec:vc} and this best-fitting model derived adopting the fiducial circular-velocity profile. We note that excluding the first two $V_{\mathrm{c}}$ points (i.e., those showing the largest discrepancies among the kinematic models, see Fig.~\ref{fig:vc}) yields best-fitting parameters consistent, within 1-$\sigma$ uncertainties, with those reported in Table~\ref{tab:dyn_bestfit}. Assuming a thick stellar disc with a scale height of 500 pc also produces best-fitting parameters consistent with the fiducial model. These mass models provide the basis for our calculation of the specific angular momentum and for the comparison of ADF22.1 with local dynamical scaling relations and their expected evolution, which we present and discuss in Sections~\ref{sec:angular} and \ref{sec:scaling}.

\begin{table} 
\begin{center}
\caption{Assumptions for the prior distributions and allowed ranges of the parameters defining the dynamical model. For parameters with truncated normal priors, the first value in the third column gives the mean of the distribution, and the second value gives the standard deviation.
}
\label{tab:tab_prior}
\begin{tabular}{ccc}
\hline\hline \noalign{\smallskip}
Parameter & Prior distribution & Range/Parameters\\
\noalign{\smallskip}
\hline
\noalign{\medskip}
$\alpha_{\mathrm{CII}}$ ($M_{\odot}/L_{\odot}$) & truncated normal & 20, 10\\
$\alpha_{\mathrm{CO}}$ ($M_{\odot}/{\mathrm{K\,km\,s^{-1}\,pc^2}}$) & truncated normal & 2.5, 1.5\\
$\alpha_{\star}$ & uniform & [0, 10]\\
$\log (M_{200}/M_{\odot})$ & uniform & [$\log(M_{\mathrm{bar}}/f_{\mathrm{bar, cosmo}})$, 14]\\
\noalign{\smallskip}
\hline
\end{tabular}
\end{center}
\end{table}

\begin{table}
\begin{center}
\caption{Best-fitting parameters of the dynamical model (upper table) and derived quantities (lower table). The molecular and diffuse gas masses are inferred from the CO(3--2) and [CII] emission lines, respectively.
}
\label{tab:dyn_bestfit}
\begin{tabular}{cc}
\hline\hline \noalign{\smallskip}
\multicolumn{2}{c}{Best-fitting parameters}\\
\hline
\noalign{\smallskip}
$\alpha_{\star}$  & 1.1$\pm$0.2
\\
$\alpha_{\mathrm{CO}}$ ($M_{\odot}/{\mathrm{K\,km\,s^{-1}\,pc^2}}$) & 1.8$^{+1.0}_{-0.9}$\\
$\alpha_{\mathrm{CII}}$ ($M_{\odot}/L_{\odot}$) & 7.7$^{+6.2}_{-5.0}$\\
$\log (M_{200}/M_{\odot})$ 
 &  12.9$^{+0.4}_{-0.3}$ \\
\noalign{\smallskip}
\hline
\noalign{\medskip}
\multicolumn{2}{c}{Derived parameters}
\\
\noalign{\smallskip}
\hline
\noalign{\smallskip}
$M_{\star}$ ($M_{\odot}$) & $2.7^{+0.5}_{-0.6} \times 10 ^ {11}$\\
$M_{\mathrm{gas, mol}}$ ($M_{\odot}$)  &  $(1.4\pm 0.8) \times 10 ^ {11}$\\
$M_{\mathrm{gas, dif}}$ ($M_{\odot}$)  &   $9.0^{+7.3}_{-5.8} \times 10 ^ {10}$ \\
$M_{\mathrm{gas}}$ ($M_{\odot}$)  & $2.5^{+0.9}_{-0.8} \times 10 ^ {11}$\\
$M_{\mathrm{b}}$ ($M_{\odot}$)  & $5.2^{+0.6}_{-0.7} \times 10 ^ {11}$\\
$\epsilon_{\star}$  & $0.22^{+0.29}_{-0.14}$\\
$\epsilon_{\mathrm{b}}$ & $0.43^{+0.61}_{-0.27}$\\
$R_{\mathrm{e, m}}$ (kpc) & $5.5 \pm 1.0$\\
$R_{200}/R_{\mathrm{e, m}}$ & $27^{+11}_{-6}$\\
\hline
\end{tabular}
\end{center}
\end{table}

\begin{figure} 
    \begin{center}  \includegraphics[width=\columnwidth]{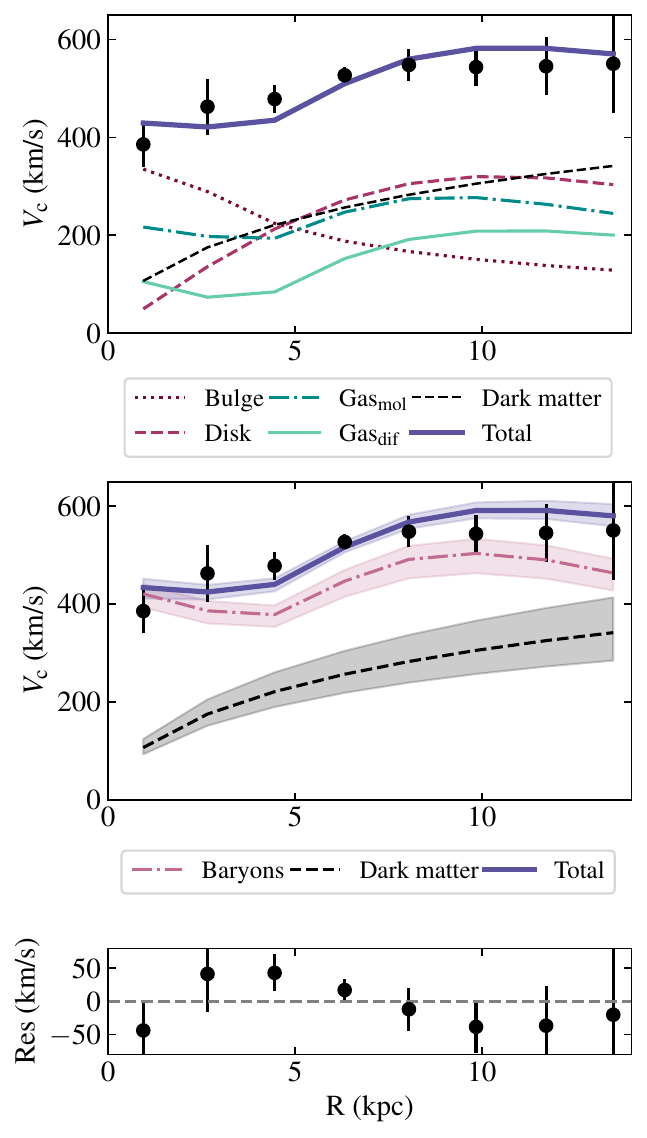}
        \caption{Circular-velocity profile of ADF22.1 and best-fitting dynamical model. The top panel shows the individual components: stars split into bulge and disc, gas split into the molecular and diffuse phases, and the dark-matter contribution as indicated in the legend. The middle panel shows the same best-fitting model, but with the stellar and gas components combined into a single baryonic contribution. The bottom panel shows the residuals. }          \label{fig:decomposition_fiducial}
    \end{center}
\end{figure}

\section{Computing the specific angular momentum}
\label{sec:angular}
In this section, we compute the specific angular momentum of the gas, stellar disc and dark matter components. For a disc, the specific angular momentum $j_{\mathrm{i}}$ of a given baryonic component $\mathrm{i}$ (i.e., stars or gas) inside a radius $R$ is given by:
\begin{equation}
    j_{\mathrm{i}} ( <R) = \frac{\int_0^R x^2 \Sigma_{\mathrm{i}}(x) V_{\mathrm{rot, i}}(x)\,\mathrm{d}x}{\int_0^R  x \Sigma_{\mathrm{i}}(x)\,\mathrm{d}x}, 
\end{equation}
where $\Sigma_{\mathrm{i}}$ and $V_{\mathrm{rot, i}}$ are the mass surface density and rotation velocity of component $\mathrm{i}$-th \citep{Romanowsky_2012}, respectively. To compute the specfic angular momentum of the stellar and gas components ($j_\star$, $j_{\mathrm{gas}})$, we adopt the best fit functional forms for the gas and stellar surface density derived in Section~\ref{sec:analysis}.

To compute $j_\star$, we need to consider that, in principle, stars and gas do not rotate at the same speed: stars typically have lower rotation velocities due to their larger velocity dispersions. However, stellar velocity measurements are unavailable for high-$z$ galaxies, and this effect is expected to be negligible for massive systems \citep{ManceraPina_2025b}. For this reason, we adopt the approximation  $V_{\mathrm{rot, \star}} = V_{\mathrm{rot, gas}}$ to avoid introducing additional systematics. To assess the impact of this assumption, we also compute $j_{\star}$ following the prescription adopted by \citet{Marasco_2019} for a sample of galaxies at $z \approx 1$, and find that the resulting values differ by only 2\%, consistent with the analysis of \citet{ManceraPina_2021}, who found that for high stellar-mass galaxies the effect of including or excluding the asymmetric-drift correction in the estimate of $j_{\star}$ is negligible. 

For computing the gas specific angular momentum, $j_{\mathrm{gas}}$, we combine the surface mass densities of the molecular and diffuse components and assume the [CII] rotation velocity. This co-rotation assumption is well justified by the agreement between the [CII] and CO(3-2) rotation velocities, derived in Section \ref{sec:kinematic} and  \citet{Rizzo_2023}, respectively.

The specific angular momentum of the baryons is then given by
\begin{equation}
  j_{\mathrm{b}} = f_{\mathrm{gas}} j_{\mathrm{gas}} + (1 - f_{\mathrm{gas}}) j_{\mathrm{\star}},
\end{equation}

where $f_{\mathrm{gas}} = M_{\mathrm{gas}}/M_{\mathrm{bar}}$. 

\begin{figure}[th!] 
    \begin{center}  \includegraphics[width=0.8\columnwidth]{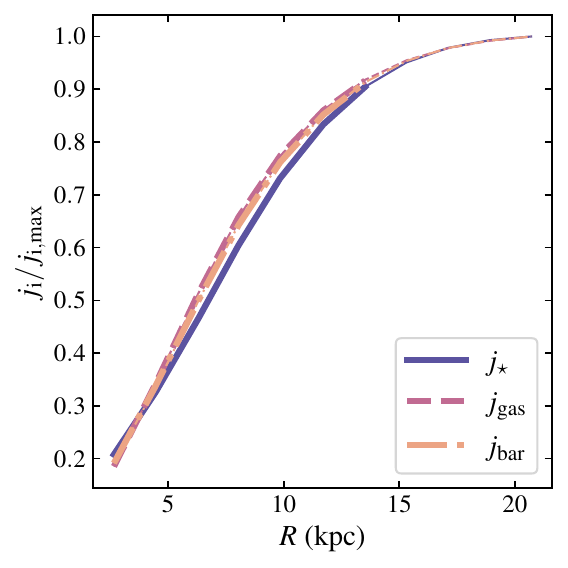}
        \caption{Cumulative specific angular momentum for the stellar, gas and baryonic components, as indicated in the legend. The thick curves show the measured $j$ and the thin curves show the extrapolated counterparts.}  
        \label{fig:jcum}
    \end{center}
\end{figure}

A robust estimate of the total specific angular momentum requires integrating to large radii, where $j$ approaches its asymptotic value. This is necessary because a substantial fraction of the angular momentum of a disc resides in its outskirts \citep{Romanowsky_2012}. To capture this contribution, we compute $j$ for stars, gas, and baryons under the assumption of a flat rotation curve, extending the integration to $R = 20$ kpc, beyond the last measured point. Observational evidence supports this assumption: \citet{Umehata_2025}, for instance, found that the rotation curve of ADF22.1 remains flat out to at least 16 kpc using tapered imaging to trace extended emission. For a conservative lower limit, we also calculate $j$ by truncating the disc at the outermost observed point of the rotation curve ($R \approx 14$ kpc), setting the surface density to zero beyond this radius and we report the values in Table \ref{tab:j}. Fig.~\ref{fig:jcum} shows the cumulative specific angular momentum profiles: the thick curve corresponds to the direct measurement up to the last observed point, while the thin curves illustrate the extrapolations. In all cases, the measured values of $j$ are above 80\% of the extrapolated ones (see Table \ref{tab:j}). As mentioned in the Introduction, the value of $j_{\star}$ was previously estimated by \citet{Umehata_2025}, and our extrapolated value is consistent with theirs within the uncertainties. \\
In addition to the specific angular momentum of the baryonic component, we estimate the specific angular momentum of the dark-matter halo, $j_{\mathrm{DM}}$. Following the definition by \citet{Bullock_2001}, this can be written as
\begin{equation}
j_{\mathrm{DM}} = \lambda \sqrt{2} R_{200} V_{200},
\end{equation}
where $\lambda$ is the spin parameter and $R_{200}$ is the radius of the dark-matter halo containing the mass $M_{200}$ and $V_{200}$ is the corresponding halo circular velocity at $R_{200}$. From the rotation curve decomposition, we obtain $M_{200}$, which allows us to compute $R_{200}$ and $V_{200}$. \\
To estimate $j_{\mathrm{DM}}$, we assume a value of $\lambda = 0.035$ \citep{Bullock_2001}. Numerous studies show that the spin parameters of dark-matter haloes follow a log-normal distribution, centered at $\lambda \approx 0.035$ with a scatter in $\ln(\lambda)$ of $\sim 0.5$, and that they exhibit no dependence on redshift  \citep[e.g.,][]{White_1984, Bullock_2001, Maller_2002, Danovich_2015, Lagos_2022}. In Sections \ref{sec:scaling} -- \ref{sec:formation}, we use the measurements of the specific angular momentum of the stellar, baryonic, and dark-matter components of ADF22.1 to compare it with local giant disc galaxies and to investigate the physical mechanisms that may have driven its formation.

\begin{table} 
\begin{center}
\caption{Specific angular momentum of the baryonic and dark-matter components. For the baryons, the second column lists the measured 
$j$ values, while the third column gives the values obtained after extrapolating the rotation curve out to 20 kpc. The specific angular momentum of the dark matter $j_{\mathrm{DM}}$ and the ratios between the stellar (baryonic) specific angular momentum to that of the dark-matter halo are obtained assuming a value of the spin $\lambda$ of 0.035.
}
\label{tab:j}
\begin{tabular}{c|cc}
\hline\hline \noalign{\smallskip}
Parameter & Measured & Extrapolated\\
\hline
\noalign{\smallskip}
$j_{\star}$ (kpc km s$^{-1}$) & $2815 \pm 83$ & $3110 \pm 72$\\
$j_{\mathrm{gas, mol}}$ (kpc km s$^{-1}$) & $3189_{-85}^{+78}$ & $3399_{-78}^{+73}$\\
$j_{\mathrm{gas}}$ (kpc km s$^{-1}$) & $3417_{-151}^{+176}$ & $3729_{-205}^{+221}$\\
$j_{\mathrm{bar}}$ (kpc km s$^{-1}$) & $3216_{-132}^{+139}$ & $3525_{-168}^{+162}$\\
$j_{\mathrm{DM}}$ (kpc km s$^{-1}$) & -- & $3436_{-1422}^{+2762}$\\
\hline
\multicolumn{3}{c}{Specific angular momentum ratios}\\
\hline
\noalign{\smallskip}
$f_{j_\star}$ & \multicolumn{2}{c}{$0.9_{-0.4}^{+0.6}$}\\
$f_{j_{\mathrm{b}}}$ & \multicolumn{2}{c}{$1.02_{-0.46}^{+0.73}$}\\
\hline
\end{tabular}
\end{center}
\end{table}

\section{ADF22.1 and the similarity with local giant discs} \label{sec:scaling}

\subsection{Stellar-to-halo mass relation}
\label{sec:SHMR}
The rotation curve decomposition presented in Section~\ref{sec:analysis} allows us to place constraints on the dark matter halo mass of ADF22.1. Combined with our measurements of the stellar and baryonic masses, these results allow us to investigate the stellar-to-halo and baryon-to-halo mass ratios of our target, which serve as direct probes of the efficiency with which ADF22.1 has retained its baryons, accreted gas, and converted them into stars. We discuss these ratios in the context of local and high-$z$ galaxies and predictions from models. To quantify these efficiencies we recast the stellar-to-halo and baryon-to-halo mass ratios in terms of the integrated star formation efficiency, both normalized to the cosmological baryon fraction $f_\mathrm{b,\,cosmo} = 0.16$,
\begin{equation}
    \epsilon_\star = \frac{M_\star}{f_\mathrm{b,\,cosmo}\,M_{200}},
\end{equation}
and the baryon condensation efficiency,
\begin{equation}
    \epsilon_{\mathrm{b}} = \frac{M_\mathrm{b}}{f_\mathrm{b,\,cosmo}\,M_{200}}.
\end{equation}

\begin{figure}
    \begin{center}  \includegraphics[width=\columnwidth]{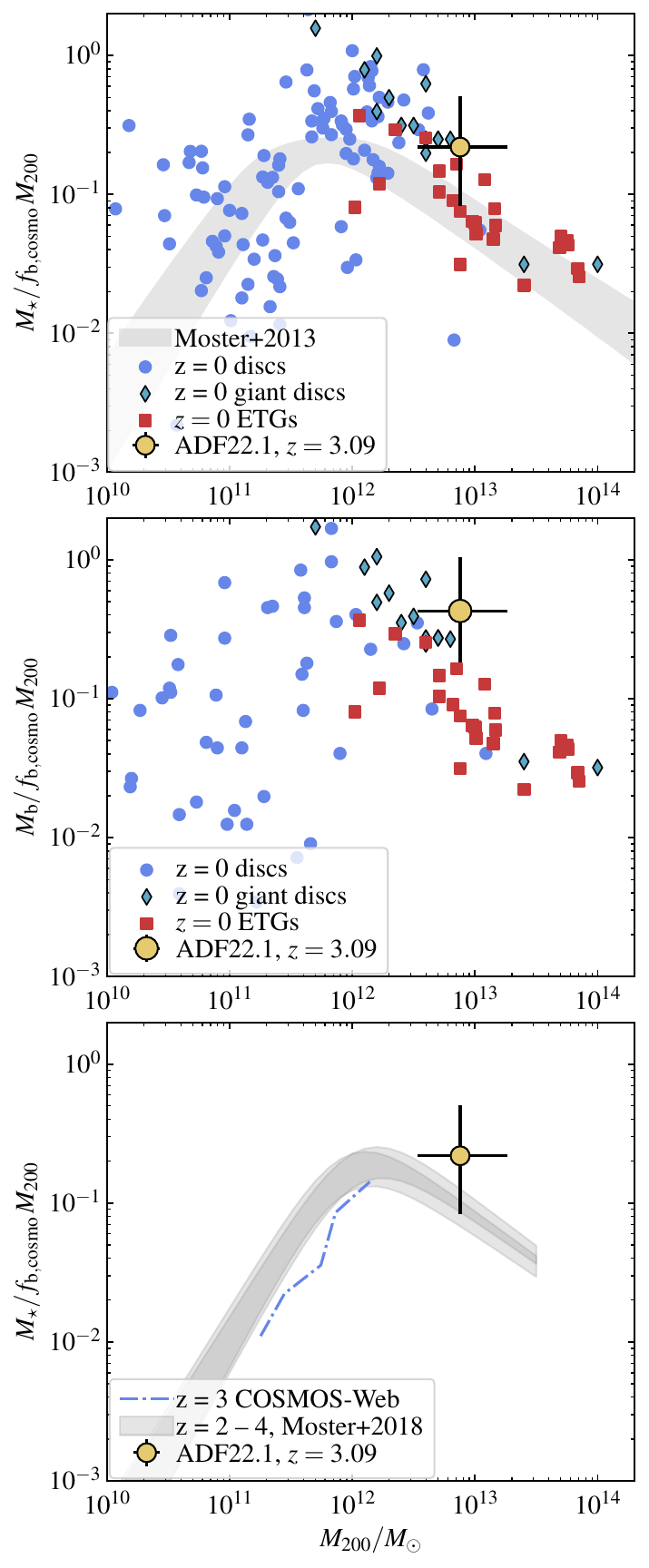}
    \vspace{-0.5cm}
        \caption{\textit{Upper panel:} Location of ADF22.1 in the stellar-to-halo mass relation in the form of the stellar-to-baryon conversion efficiency $\epsilon_{\star}$ as a function of the halo mass, compared to local normal discs \citep{Posti_2019}, local giant discs \citep{DiTeodoro_2023}, early-type galaxies \citep{Posti_2021} and models \citep{Moster_2013}. \textit{Middle panel:} same as above but for the baryon condensation efficiency $\epsilon_{\mathrm{b}}$, with $z=0$ discs from \citet{ManceraPina_2025}. \textit{Bottom panel:} Stellar efficiency of ADF22.1 compared to the expected efficiencies at $z=2$ and $z=4$ from \citet{Moster_2018} and from the conversion efficiency at $z = 3$, derived by \citet{Paquereau_2025} using the COSMOS-Web survey.}  
        \label{fig:masshalo}
    \end{center}
\end{figure}

In Table \ref{tab:dyn_bestfit}, we report the values of $\epsilon_\star$ and $\epsilon_{\mathrm{b}}$ for our target. In Fig.~\ref{fig:masshalo} (upper and middle panels), we compare the position of ADF22.1 in the $\epsilon_\star$--$M_{200}$ and $\epsilon_{\mathrm{b}}$--$M_{200}$ planes with samples of $z = 0$ galaxies. In the $\epsilon_\star$--$M_{200}$ plane, we include samples of disc and early-type galaxies at $z = 0$ \citep{Posti_2019, Posti_2021, DiTeodoro_2023}, whose halo masses are derived from dynamical modelling using methods broadly comparable to those adopted here; the disc galaxy sample includes local giant discs.

It is well established that, at fixed halo mass and $z = 0$, disc galaxies have systematically higher $\epsilon_\star$ than early-type galaxies, pointing to a more efficient conversion of baryons into stars in rotationally supported systems. The comparatively lower efficiencies of early-type galaxies are thought to reflect the combined action of mergers and AGN feedback in suppressing star formation \citep{Posti_2019, Posti_2021, Kakos_2024, Wang_Kai_2025}. Strikingly, ADF22.1 at $z \sim 3$ already reaches star formation efficiencies comparable to those of local giant disc galaxies of the same halo mass, and lies above the locus of early-type galaxies.

A consistent picture emerges from the baryon condensation efficiency plane (middle panel of Fig.~\ref{fig:masshalo}), where ADF22.1 lies in a region broadly consistent with local massive disc galaxies \citep{DiTeodoro_2023, ManceraPina_2025}, and well above the locus occupied by early-type galaxies. For the latter, we assume $M_{\mathrm{b}} \approx M_\star$, as local early-type systems typically have gas fractions below $10\%$ \citep[e.g.,][]{Serra_2012}.. Taken together, these comparisons indicate that ADF22.1 had already assembled stellar, baryonic, and dark matter halo masses comparable to those of present-day massive discs by $z \sim 3$.

In the bottom panel of Fig. \ref{fig:masshalo}, we compare the location of ADF22.1 in $\epsilon_\star$
--$M_{200}$ plane with respect to predictions at $z=2$ and $z=4$ for central star-forming galaxies from the semi-analytic model by \citet{Moster_2018}, as well as with the empirical relation inferred by \citet{Paquereau_2025} from COSMOS-Web galaxies at $z = 3$. Although the COSMOS-Web constraints do not extend to the high halo-mass regime probed by our target, they are qualitatively consistent with, and systematically lower than, the \citet{Moster_2018} relation over the mass range where the two overlap. We note that the COSMOS-Web relation is derived within a halo occupation distribution framework and therefore does not rely on dynamical halo-mass measurements as for our target. In this comparison, our target has a higher stellar-to-baryon conversion efficiency than predicted at fixed halo mass. In other words, ADF22.1 appears to have converted a larger fraction of its available baryons into stars than is typically expected for galaxies at similar stellar or halo masses. 

Given that ADF22.1 hosts the brightest X-ray AGN in the central region of the protocluster, this suggests that, at earlier epochs, AGN feedback was not sufficiently effective at removing gas to suppress star formation and therefore halting the growth of an extended disc. The current AGN activity may still influence the later stages of the galaxy evolution; however, its impact could be limited by the already deep gravitational potential well of the system. 

\subsection{Comparing the stellar and baryonic $j$ to that of the dark-metter halo}
\label{sec:fj}
As discussed in the introduction, it is well established that for disc galaxies at $z = 0$ the ratio between the specific angular momentum of the stellar (or baryonic) component and that of the dark matter halo, $f_{j_\star} = j_\star/j_{\mathrm{DM}}$ (or $f_{j_\mathrm{b}} = j_{\mathrm{b}}/j_{\mathrm{DM}}$) is close to unity and has only a weak dependence on stellar or baryonic mass \citep{Posti_2019, ManceraPina_2021, DiTeodoro_2023, Romeo_2023}. It remains unclear why this is the case, given that a number of physical processes can drive $f_j$ away from unity in either direction. These include the preferential removal of low-angular-momentum gas due to outflow, angular momentum transfer from baryons to the dark matter halo via dynamical friction,the cumulative effect of past mergers, the angular momentum stratification in the CGM \citep{Barnes_1988, Dutton_2012, Kassin_2012, Romanowsky_2012, Genel_2015, Zavala_2016, Lagos_2018}.

\begin{figure*}
    \begin{center}  \includegraphics[width=\textwidth]{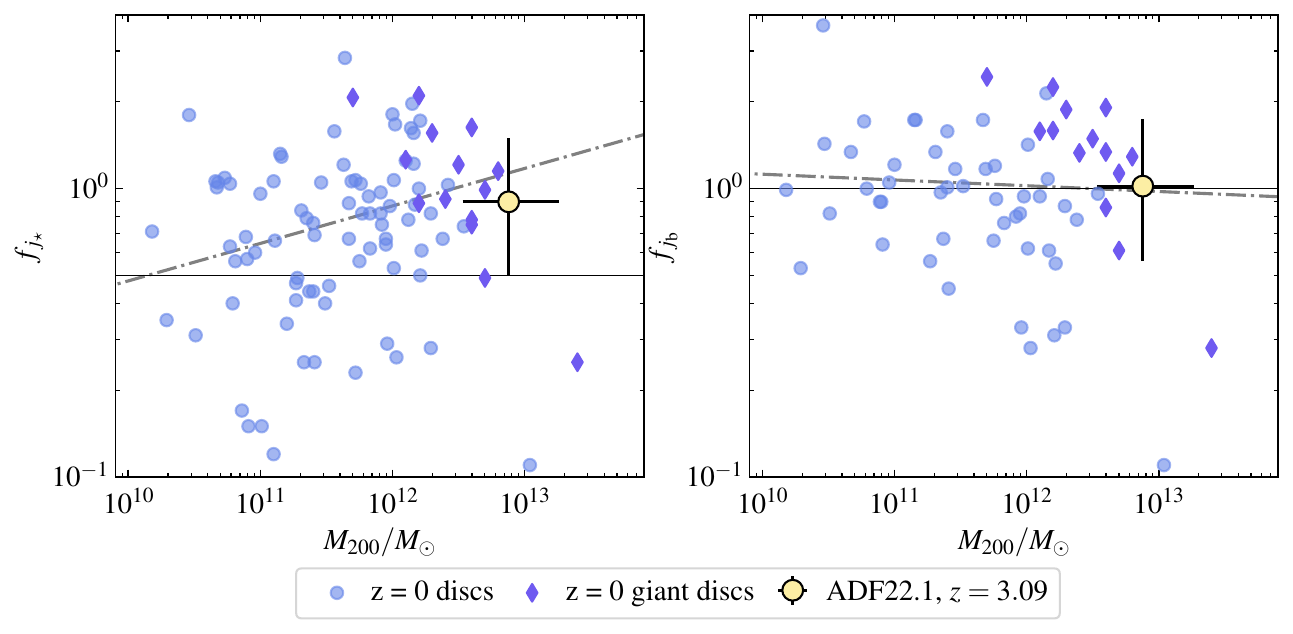}
    \vspace{-0.5cm}
        \caption{Location of ADF22.1 in the $f_{j_\star}$--$M_{200}$ (left) and $f_{j_\mathrm{b}}$--$M_{200}$ (right) planes, compared to local discs \citep[circles;][]{Posti_2019, ManceraPina_2021} and local giant discs \citep[diamonds;][]{DiTeodoro_2023}. The dot-dashed lines show the best-fitting relations from \citet{DiTeodoro_2023}. The soild horizontal lines indicate the indicative values of $f_{j_\star} = 0.5$ and $f_{j_\mathrm{b}} = 1$, typically reported in simulations \citep[e.g.,][]{Danovich_2015, Teklu_2015, Lagos_2017}.}  
        \label{fig:fj}
    \end{center}
\end{figure*}

In this section, we compare $f_{j_\star}$ and $f_{j_\mathrm{b}}$ of ADF22.1 with those of local galaxies, including giant discs. Fig.~\ref{fig:fj} shows the distribution of both quantities as a function of $M_{200}$, together with the best-fitting local relation from \citet[][dot-dashed line]{DiTeodoro_2023}. The values of $f_{j_\star}$ and $f_{j_\mathrm{b}}$ for ADF22.1, also reported in Table~\ref{tab:j}, are $0.9_{-0.4}^{+0.6}$ and $1.0_{-0.5}^{+0.7}$, respectively. In both the $f_{j_\star}$--$M_{200}$ and $f_{j_\mathrm{b}}$--$M_{200}$ planes, ADF22.1 is indistinguishable from local discs, confirming that, despite being at $z \sim 3$, it is structurally equivalent to the most massive local discs. For reference, in Fig.~\ref{fig:fj}, we overlay with thin solid lines the indicative values $f_{j_\star} = 0.5$ and $f_{j_\mathrm{b}} = 1$, as typically reported in simulations \citep[e.g.,][]{Danovich_2015, Teklu_2015, Lagos_2017}, which find that $j_\star \sim j_\mathrm{DM}$ to within a factor of two, and that the baryonic specific angular momentum exceeds its stellar counterpart on average. The values of $f_{j_\star}$ and $f_{j_\mathrm{b}}$ measured for ADF22.1 are in good agreement with these predictions, within the uncertainties.

\subsection{Dynamical scaling relations}
\label{sec:dyn_scal}
\begin{figure*}
    \begin{center}  \includegraphics[width=\textwidth]{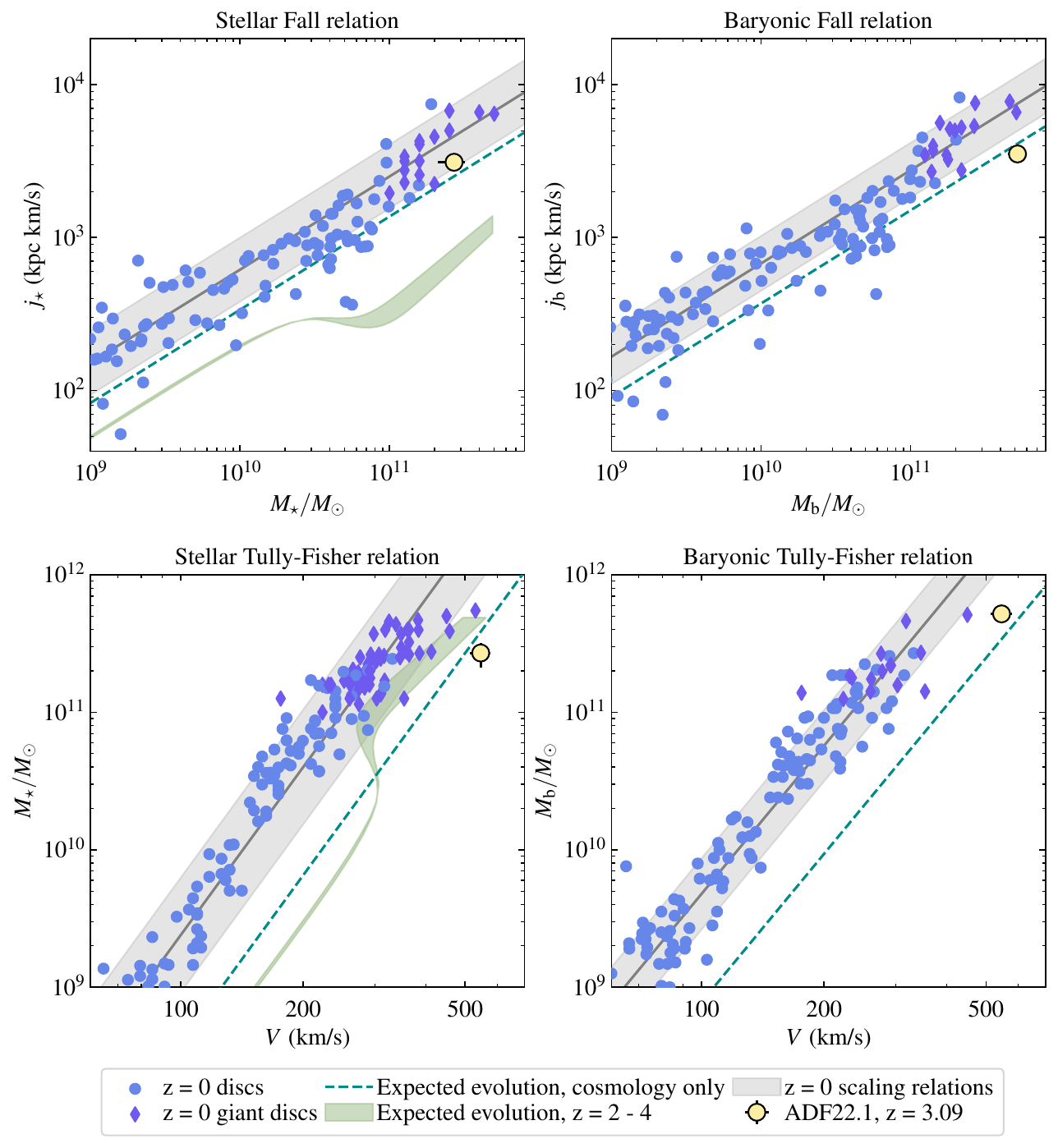}
        \caption{Location of ADF22.1 (circles) in the Fall relation (top panels) and the Tully-Fisher relation (bottom panels). We compare ADF22.1 to local normal discs (circles) and local giant discs (thin diamonds), together with the $z=0$ best-fitting relations and their scatter (solid lines and corresponding shaded regions). We also show the expected redshift evolution of these relations: (i) the evolution only driven by changes in critical overdensity and the Hubble parameter (dashed lines), and (ii) the evolution predicted when including cosmological effects, the redshift dependence of the stellar-to-halo mass relation from \citet{Moster_2018}, and the evolution of the stellar-to-halo specific angular momentum. For the latter, we plot the expected evolution over $z=2-4$ (green shaded regions). The $z=0$ disc samples are taken from \citet{ManceraPina_2021} for the Fall relation and from \citet{Lelli_2016, Lelli_2019} for the Tully--Fisher relation, while the local giant-disc data are from \citet{DiTeodoro_2021, DiTeodoro_2023}. The $z=0$ relations are from \citet{DiTeodoro_2023}.}  
        \label{fig:scalingrelations}
    \end{center}
\end{figure*}

In this section, we use the stellar and baryonic Fall and Tully--Fisher relations at $z=0$ as a reference to derive the expected redshift evolution of these scaling relations and to compare them with the location of ADF22.1. Fig.~\ref{fig:scalingrelations} shows the best-fitting $z=0$ stellar and baryonic Tully--Fisher and Fall relations from \citet{DiTeodoro_2023}, together with their intrinsic scatter (gray solid lines and shaded regions). The blue circles and purple thin diamonds indicate the positions of local discs and giant discs compiled from the literature \citep{Lelli_2016, Lelli_2019, ManceraPina_2021, DiTeodoro_2023}, while the yellow circle marks the location of ADF22.1. 

Our target lies on the local stellar Fall relation \citep[see also ][]{Umehata_2025}, but falls below the $z=0$ baryonic Fall relation as well as both the stellar and baryonic Tully--Fisher relations. To understand the relative position between ADF22.1 and local relations, we note that in ADF22.1 the gas and stellar components have similar mass and specific angular momentum: $M_{\mathrm{b}} \approx 1.9\,M_\star$ and $j_{\mathrm{b}} \approx 1.1\,j_\star$. As a result, the position of ADF22.1 in the baryonic scaling relations is well approximated by a horizontal shift of $\approx 0.3$~dex with respect to its position in the stellar relations. In contrast, at $z = 0$, high-mass disc galaxies are typically gas-poor, such that $M_{\mathrm{b}} \approx M_\star$ and $j_{\mathrm{b}} \approx j_\star$, with no appreciable horizontal offset between the stellar 
and baryonic relations. 

For completeness, we note that the total gas specific angular momentum of ADF22.1 lies a factor 
of $\approx 30$ below the local $j_{\mathrm{gas}}$--$M_{\mathrm{gas}}$ relation \citep{ManceraPina_2021}. However, this comparison should be interpreted with caution, as the local relation is an extrapolation at the gas masses relevant for ADF22.1: no local galaxy has 
$M_{\mathrm{gas}} \gtrsim 10^{11}\,\mathrm{M}_\odot$. More fundamentally, the large gas reservoirs present at $z \approx 3$ are expected to have been largely converted into stars by $z = 0$. Furthermore, while the molecular specific angular momentum $j_{\mathrm{gas,mol}}$ 
of ADF22.1 is consistent within the uncertainties with the local molecular $j$--$M$ relation \citep{Geesink_2025}, a direct comparison with the total-gas relation in nearby galaxies is not straightforward: at low redshift, the gas mass is dominated by neutral atomic gas traced by 
HI, whereas the atomic-to-molecular ratio at $z \approx 3$ remains uncertain \citep{Tacconi_2020}.

To interpret the position of ADF22.1 in the four planes shown in Fig.~\ref{fig:scalingrelations}, we consider how the Fall and Tully–Fisher relations are expected to evolve from $z=0$ to $z=3$. Following the approach of \citet{ManceraPina_2025b}, we begin by writing the virial mass--velocity and mass--angular-momentum relations for CDM haloes:

\begin{equation}
M_{\mathrm{vir}} =
\frac{1}{G\,H(z)}
\sqrt{\frac{2}{\Delta_c(z)}} \,
V_{\mathrm{vir}}^{3},
\label{eq:mvir}
\end{equation}

\begin{equation}
j_{\mathrm{vir}} = \lambda
\frac{\left( 2\,G M_{\mathrm{vir}}\right)^{2/3} }
{[H(z)]^{1/3} [\Delta_c(z)]^{1/6}}
\label{eq:jvir}
.
\end{equation}

where $G$ is the gravitational constant, $\Delta_c(z)$ is the critical overdensity for virialisation, $H(z)$ is the Hubble parameter, and $\lambda$ is the halo spin parameter (see Section \ref{sec:angular}). These relations assume isothermal haloes and are accurate to within $\sim$10\% for haloes with concentrations of $c \simeq 3$–4 \citep{Bullock_2001}. If we express equation (\ref{eq:mvir}) and (\ref{eq:jvir}) in terms of observable quantities (i.e., $M_\star$, $V$, $j_\star$) and the ratios between stellar and dark-matter halo properties $M_{\mathrm{vir}}$, $V_{\mathrm{vir}}$, $j_{\mathrm{DM}}$ ($f_{M_\star} = M_\star/M_{\mathrm{vir}} = x \epsilon_{\star} f_{\mathrm{b, cosmo}}$, where $x$ is the conversion factor between the virial mass and $M_{200}$; $f_V = V/V_{\mathrm{vir}}$; $f_{j_\star}$), we obtain:

\begin{equation}
M_\star =
\sqrt{2}
\left( \frac{f_{M_\star}}{f_V^{3}} \right)
\frac{1}{G\,H\,\sqrt{\Delta_c}}
\, V^{3} ,
\end{equation}

\begin{equation}
j_\star = \frac{\lambda f_{j_\star}}{[H(z)]^{1/3} [\Delta_c(z)]^{1/6}}
\left( 
\frac{2\,G \,M_{\mathrm{\star}}}{f_{M_\star}} 
\right)^{2/3} 
\end{equation}

Evaluating these equations at redshift $z$ and at $z=0$, at fixed rotation velocity and stellar mass, respectively, yields:

\begin{equation}
\frac{M_\star(V,z)}%
     {M_\star(V,0)}
=
\left[
\frac{f_{M_\star}(M_\star,z)}{f_{M_\star}(M_\star,0)}
\right]
\left[
\frac{f_V(M_\star,z)}{f_V(M_\star,0)}
\right]^{-3}
\left[
\frac{\Delta_c(z)}{\Delta_c(0)}
\right]^{-1/2}
\left[
\frac{H(z)}{H(0)}
\right]^{-1},
\label{eq:evtfr}
\end{equation}

\begin{equation}
\frac{j_\star(M_\star,z)}%
     {j_\star(M_\star,0)}
=
\left[
\frac{f_{j_\star}(M_\star,z)}{f_{j_\star}(M_\star,0)}
\right]
\left[
\frac{f_{M_\star}(M_\star,z)}{f_{M_\star}(M_\star,0)}
\right]^{-2/3}
\left[
\frac{\Delta_c(z)}{\Delta_c(0)}
\right]^{-1/6}
\left[
\frac{H(z)}{H(0)}
\right]^{-1/3}.
\label{eq:evfall}
\end{equation}

Analogous expressions apply to the baryonic versions of these relations. Following \citet{ManceraPina_2025b}, in Fig.~\ref{fig:scalingrelations} we plot the evolutionary tracks at $z=3.09$ expected from cosmology alone (green dashed lines), i.e. assuming that $f_{M_\star}$, $f_{j_\star}$, and $f_V$ do not evolve with redshift, though they may depend on mass. In other words, we consider the case in which the first two terms on the right-hand side of equations (\ref{eq:evtfr}) and  (\ref{eq:evfall}) are unity, and the tracks are set entirely by the change in the critical overdensity and the Hubble parameter: $\Delta_{\rm c}=101$ and 174 \citep{Bryan_1998}, and $H(z)=66.7$ and $317\ {\rm km\ s^{-1}\ Mpc^{-1}}$ at $z=0$ and $z=3.09$, respectively. The results described in Section~\ref{sec:SHMR} indicate that the star-formation and baryon conversion efficiencies $\epsilon_\star$ and $\epsilon_\mathrm{b}$ of ADF22.1 are consistent with those of local discs. Combined with the fact that $x(z=0)/x(z=3.09) \approx 1$, this confirms that the assumption of a non-evolving
$f_{M_\star}$ is well justified for ADF22.1. Similarly, the results reported in Section~\ref{sec:fj}, together with the measured velocity ratio $f_V = 0.8^{+0.3}_{-0.1}$, consistent with values found for local discs \citep{Posti_2019, ManceraPina_2025}, confirm that the assumptions of non-evolving $f_{j_\star}$ and $f_V$ are likewise well justified for this galaxy.

The observed position of ADF22.1 in both the Tully–Fisher and Fall planes is consistent with the cosmology-only evolution of the scaling relations, assuming they have a similar scatter to their local counterparts. This confirms that ADF22.1 is structurally equivalent to the giant discs observed at $z=0$, with its apparent offset from the local relations driven primarily by the cosmological rescaling of $\Delta_{\rm c}$ and $H(z)$ at high redshift, and not from an evolution in the intrinsic disc and halo properties.

As a reference, in Fig.~\ref{fig:scalingrelations}, we also show the expected evolution of the stellar Tully–Fisher and Fall relations under the assumption that $f_V$ remains constant with redshift, while $f_{j_{\star}}$ and $f_{M_{\star}}$ evolve (shaded green regions). For $f_{j_{\star}}$, we adopt $0.9$ at $z = 0$, consistent with measurements for local disc galaxies \citep{DiTeodoro_2023, ManceraPina_2021}, and $f_j = 0.5$ at high-$z$ (see references in Section \ref{sec:fj}). For $f_{M_{\star}}$, we assume the same semi-analytical model by \citet{Moster_2018} described in Section~\ref{sec:SHMR}. Compared to the cosmology-only scenario, allowing $f_{j_{\star}}$ and $f_{M_{\star}}$ to evolve shifts the Fall relation toward lower specific angular momentum at fixed stellar mass, which are largely inconsistent with our measurement, while the stellar Tully–Fisher relation remains broadly consistent with the cosmology-only prediction.

\section{A giant spiral and its descendants}
\label{sec:descend}
\subsection{Identifying the descendants}
\citet{Wang_2025} and \citet{Umehata_2025} proposed that the $z=3$ giant discs may evolve into the progenitors of $z=0$ most massive elliptical galaxies in the cores of galaxy clusters. However, no quantitative comparison has yet been carried out to test this connection. In this section, we investigate the potential $z=0$ descendants of ADF22.1. To this end, we use the Mapping Nearby Galaxies at Apache Point Observatory (MaNGA) survey \citep[SDSS DR17;][]{Abdurrouf_2022}, which comprises a sample of $\sim10{,}000$ nearby galaxies with publicly available measurements of their kinematics, stellar populations, and environmental properties \citep[e.g.,][]{Tempel_2014, Tinker_2021, Sanchez_2022, Zhu_2023}.

We identify potential descendants by applying the following selection criteria to 
the MaNGA sample:

\begin{itemize}

\item Circular velocity and stellar mass. We retain only galaxies with $V_{\mathrm{c}} > 514~\mathrm{km\,s^{-1}}$ and $\log(M_\star/M_\odot) > 11.32$. Because the gravitational potential of a galaxy is expected to deepen over cosmic 
time as mass increases, potential descendants of ADF22.1 should have circular velocities and stellar masses at least as large as those of the progenitor. Circular velocity is particularly well suited for this purpose as it serves as a direct proxy 
for the depth of the gravitational potential well. To remain conservative, we set both thresholds at the best-fitting value of ADF22.1 minus $1\sigma$. For the MaNGA galaxies, circular velocities are taken from the DynPop project \citep{Zhu_2023, 
Zhu_2025}, which derives these quantities via Jeans Anisotropic Modelling \citep{Cappellari_2008, Cappellari_2020} and their stellar masses are from \citep{Sanchez_2022}.

\item Environment: no isolated galaxies. Since ADF22.1 resides in a dense protocluster environment, its descendants are unlikely to be isolated at $z=0$. We therefore remove isolated galaxies, defined as central galaxies with no satellites, and retain only galaxies classified as centrals 
($P_{\mathrm{sat}} < 0.1$) or satellites ($P_{\mathrm{sat}} \ge 0.9$). Following \citet{Angeloudi_2024}, we use the self-calibrating halo-based group catalogue of \citet{Tinker_2021}, which provides for each galaxy the satellite 
probability $P_{\mathrm{sat}}$ and, for central galaxies, the number of satellites within the host halo $N_{\mathrm{sat}}$.

\item Halo mass. We select only galaxies which are part of a group or a cluster with halo masses $\geq 5 \times 10^{13}\,\mathrm{M}_\odot$. This is a conservative threshold given that SSA22, the protocluster hosting ADF22.1, is expected to evolve into a cluster of mass $\gtrsim 10^{14}\,\mathrm{M}_\odot$  by $z=0$ \citep{Kubo_2016}; this estimate is further supported by \citet{Chiang_2013}, whose modelling of protoclusters predicts a consistent present-day halo mass when adopting the LBG overdensity $\delta_{\rm LBG} = 5.0 \pm 1.2$ measured in SSA22 by \citet{Steidel_2000}.  The values of the halo masses are from the catalogue of \citet{Tinker_2021}. 

\end{itemize}

After these selections, the descendant sample contains 27 galaxies. To assess whether these potential descendants have properties distinct from those of the general high-mass galaxy population, we construct a comparison (``full'') sample by applying all of the above criteria except the cut in circular velocity. This full sample consists of 397 galaxies, of which the 27 descendants represent $\sim7\%$. In the following subsections, we examine potential systematic differences between the two samples in terms of environment, morphology, and stellar kinematics.

\subsubsection{Environment and halo mass}

\begin{figure*}
    \begin{center}  \includegraphics[width=0.8\textwidth]{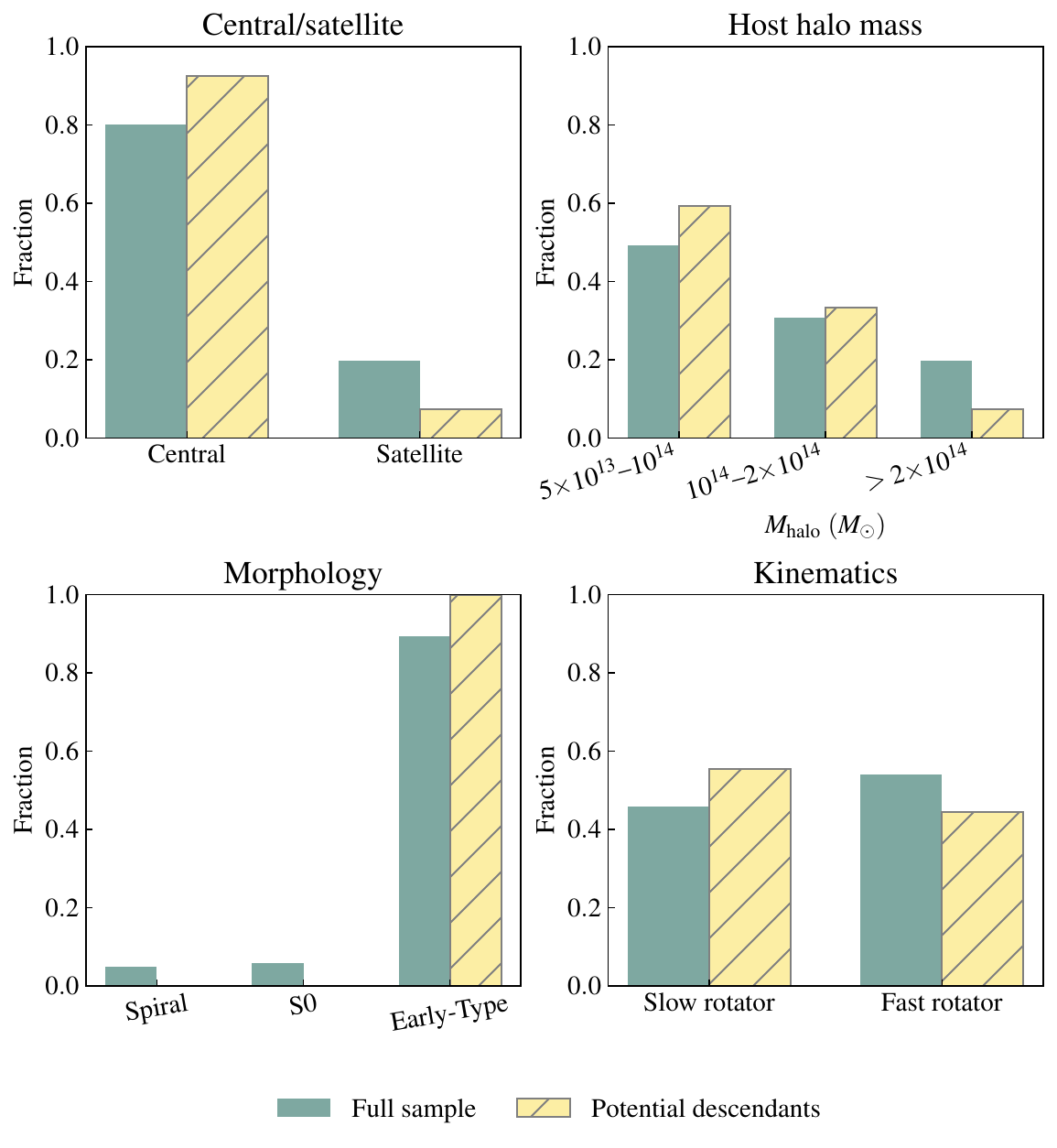}
        \caption{Comparison of candidate descendant galaxies of ADF22.1 with the full MaNGA massive-galaxy sample (see legend). \textit{Upper panels:} Fraction of galaxies classified as centrals or satellites (left), and mass distributions of their host group or cluster halo (right). \textit{Lower panels:} Distributions by morphological classes (left), and stellar kinematic classes (right).        
        }  
        \label{fig:environm}
    \end{center}
\end{figure*}

In the upper panels of Fig.~\ref{fig:environm}, we compare the environmental properties of the full (green) and descendant (yellow hatched) samples. The left panel shows the central versus satellite classification, while the right panel shows the distribution of host halo masses across three bins. The two samples inhabit broadly similar environments: the fractions of central galaxies are $80 \pm 6\%$ and $93 \pm 25\%$ for the full and descendant samples, respectively, and the satellite fractions are correspondingly consistent. The halo mass distributions of the two samples are consistent within $1\sigma$, although the descendant sample shows a mild preference for lower halo masses (i.e., $< 10^{14} M_{\odot}$; $59 \pm 19\%$ for the descendant sample, compared to $49 \pm 4\%$ for the full sample). While not statistically significant, this trend suggests that, relative to the full sample, the descendant selection preferentially yields central galaxies hosted in $M_{200} < 10^{14}\,\mathrm{M}_\odot$ haloes, rather than satellite galaxies or members of more massive clusters. This is consistent with the expectation that the most massive overdensities at early cosmic times do not necessarily evolve into the most massive structures at $z=0$: the growth of some early-collapsing systems can be suppressed by the competition between gravitational collapse and cosmic expansion \citep{Remus_2023}.

\subsubsection{Morphology and stellar kinematics}
In the bottom panels of Fig.~\ref{fig:environm}, we compare the morphological (left panel) and kinematic (right panel) classifications of the full and descendant samples. Morphologies are assigned using the visual classification classes of spirals, S0s, and ellipticals provided in the \textsc{pyMorph} catalogue \citep{Dominguez_2022}.  The full sample contains $5\pm 1\%$, $6 \pm 1\%$, and $89 \pm 7\%$ spirals, S0s, and ellipticals, respectively, while the descendant sample consists entirely of ellipticals ($100\%$). However, this apparent difference is consistent with low-number statistics. Assuming that the descendant sample is drawn from the same morphological distribution as the full sample, we expect $\sim$$1.3$ spirals among the 27 descendants, fully consistent with the observed value of zero within Poissonian uncertainties. The same argument applies to S0s.  We therefore conclude that the two samples are morphologically indistinguishable, and that the potential descendants of ADF22.1 are not preferentially drawn from any particular morphological class within the high-mass galaxy population.

Morphology alone, however, is a coarse discriminant. A more quantitative characterisation is provided by the stellar kinematic classification into fast and slow rotators, which is based on the relative importance of ordered rotation versus random motions. This is quantified through the stellar spin parameter $\lambda_{R_e}$, a proxy for the degree of rotation support \citep{Emsellem_2007, Emsellem_2011}, combined with the projected ellipticity \citep{Cappellari_2026}. The full sample comprises $54 \pm 5\%$ fast rotators and $46 \pm 4\%$ slow rotators, whereas the descendant sample is slightly more skewed towards slow rotators, with $44 \pm 15\%$ fast rotators and $56 \pm 18\%$ slow rotators. Although the two samples are consistent within the uncertainties, there is a modest but systematic offset: the descendant candidates are preferentially slow rotators, while the full sample is 
dominated by fast rotators. Interpreted in the context of the future evolution of ADF22.1, these results suggest that transformation into a spheroidal, slow-rotator system is the more likely outcome (see further discussion in Section~\ref{sec:masssize}), although evolution into a fast-rotating early-type galaxy, with a surviving disc component, cannot be ruled out.

\subsection{Mass-size comparison}
\label{sec:masssize}
As discussed in the Introduction, the location of ADF22.1 in the mass-size plane at $z = 3$ has been studied recently by \citet{Umehata_2025}, and similarly to another massive spiral galaxy at a similar redshift (i.e., the Big Wheel), it was found that they are outliers from the expected extrapolation of the stellar mass–size relation \citep{Wang_2025}. In this section, we further place ADF22.1 in the context of the local galaxy population, comparing its location in the mass–size plane to both the $z=0$ descendant candidates identified in the previous section and the population of high-$z$ brightest group galaxies (BGG) recently studied by \citet{Gozaliasl_2025}.

\begin{figure*} 
    \begin{center}  \includegraphics[width=\textwidth]{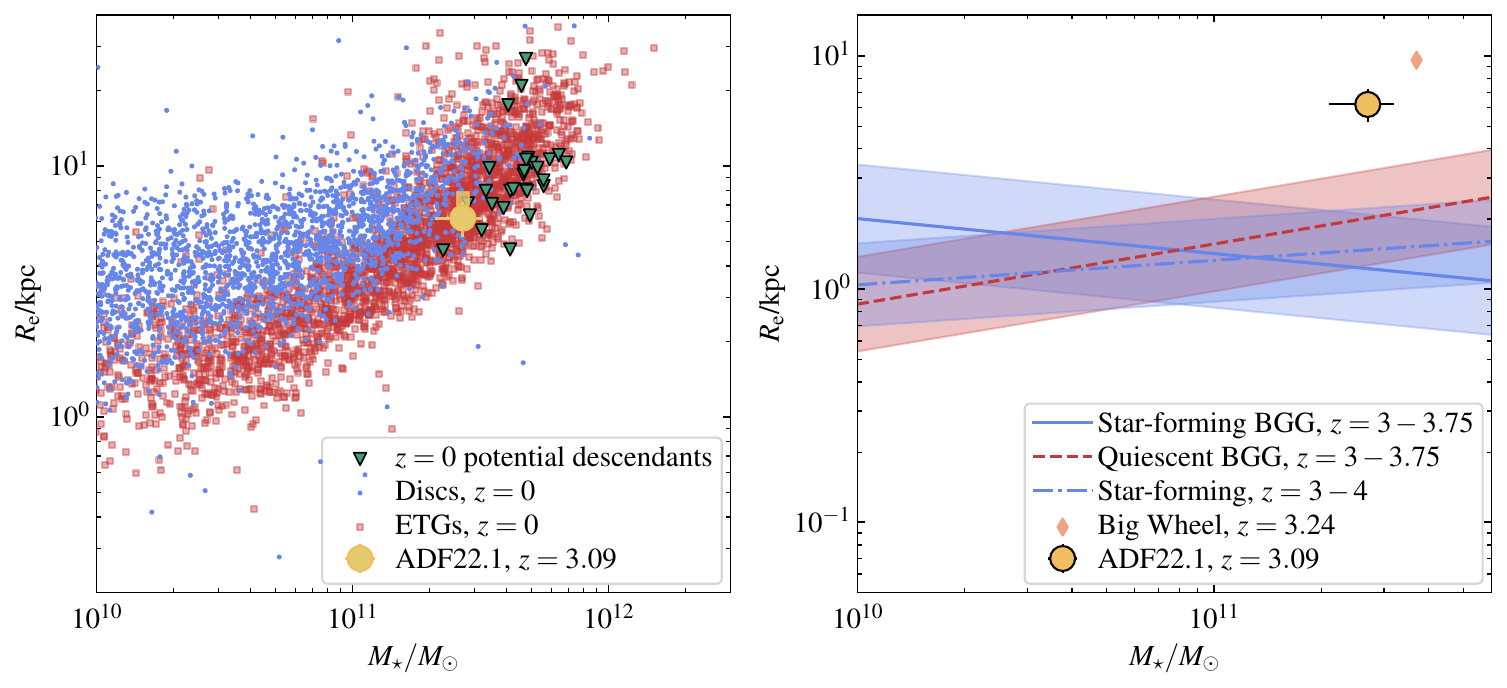}
        \caption{Location of ADF22.1 (yellow circle) in the stellar mass--size plane compared with $z = 0$ galaxies (left panel) and high-redshift samples (right panel). \textit{Left panel:} MaNGA massive galaxies split into early-type galaxies (ETGs: S0s and ellipticals; red squares) and discs (blue circles), together with the ADF22.1 descendant candidates (green triangles). The yellow square shows the location of ADF22.1 after increasing its size by 20\% to account for the systematic offset between sizes measured in the $r$-band (used for the local sample) and at rest-frame $1\,\mu$m (used for ADF22.1). \textit{Right panel:} extrapolated best-fitting relations and intrinsic scatter from \citet{Allen_2025} for star-forming galaxies at $z = 3$--$4$, and from \citet{Gozaliasl_2025} for star-forming and quiescent brightest group galaxies at $z = 3$--$3.75$. The green diamond marks the location of the Big Wheel \citep{Wang_2025}.
        }  
        \label{fig:masssize}
    \end{center}
\end{figure*}

ADF22.1 has a half-mass radius $R_{\mathrm{e, m}}$ of $5.5 \pm 1.0$ kpc, consistent within the uncertainties with its half-light radius a rest-frame wavelength of $\sim 1\mu$m of $R_{\mathrm{e}} = 6.2 \pm 0.1$ kpc. Because the quoted uncertainty on $R_{\mathrm{e}}$ is much smaller than the difference between the two estimates, and to conservatively account for potential systematics, we adopt an uncertainty of 1 kpc on the half-light radius. This is reflected in Fig. \ref{fig:masssize} by the yellow circles.

The left panel of Fig.~\ref{fig:masssize} shows the stellar mass--size distribution of MaNGA galaxies, separated into disc (blue circles) and early-type (red squares) systems. The disc class comprises the spirals, while early types include S0s and ellipticals. The descendant candidates are shown as green triangles, with a median stellar mass of $4 \times 10^{11}\,M_{\odot}$ and a median effective radius of $9$ kpc. Since MaNGA sizes are derived from $r$-band surface brightness fitting, they are expected to be $\sim$20\% larger than sizes measured at $1\,\mu$m \citep[e.g.,][]{Kelvin_2012, Vulcani_2014}, the rest-frame wavelength probed by F444W at the redshift of ADF22.1. To account for this, we also show ADF22.1 as a yellow square whose size has been increased by 20\%. Dividing the full sample into two stellar mass bins centred at $\log(M_\star/M_\odot) = 11.4 \pm 0.2$ and $11.8 \pm 0.2$, we find median effective radii (with 16th--84th percentile ranges) of $9^{+4}_{-3}$ kpc and $13^{+6}_{-4}$ kpc, respectively. The corresponding values for the descendant candidates are $7\pm1$ kpc and $10^{+1}_{-2}$ kpc, systematically smaller, despite within the scatter, indicating that the descendants preferentially occupy the compact end of the mass--size relation rather than representing extreme outliers. A direct comparison of median properties further highlights this compactness: the stellar mass and effective radius of the descendant candidates exceed those of ADF22.1 by only factors of 1.5 and 1.4, respectively. This comparison is conservative, as the true size offset may be even smaller: once the potential overestimation of descendant sizes, arising from the use of optical-wavelength imaging, is corrected for, the size ratio reduces to $\sim$1.1 (see yellow square).

In the right panel of Fig. \ref{fig:masssize}, we compare the position of ADF22.1 in the stellar mass–size plane with empirical mass–size relations derived from \textit{JWST} data at similar redshift and derived from surface brightness fitting of F444W images. We note that, unlike the left panel, this comparison does not separate galaxies by morphology based on visual classification; instead, the mass--size relations are defined by the standard star-forming/quiescent dichotomy. While the star-forming population at these redshifts is expected to be dominated by disc-like morphologies \citep[e.g.,][]{Kartaltepe_2023}, we caution that the relations may be subject to biases introduced by the irregular morphologies, more prevalent among high-$z$ galaxies.

Because ADF22.1 is the brightest galaxy in the central region of the protocluster complex, we first compare it to the relations found by \citet{Gozaliasl_2025} for the star-forming and quiescent BGGs (solid blue and dashed red lines). We plot their best-fit relation in the redshift interval $3.0<z\leq3.75$. We note that these relations are poorly constrained at high stellar masses, as only four galaxies in their sample have $\log{M_\star/M_{\odot}} \gtrsim 11.2$. Based on comparisons of massive galaxies over $0\lesssim z\lesssim 3.7$, those authors conclude that BGGs are systematically more compact than field galaxies at fixed stellar mass and redshift. As a reference, we also include the mass–size relation from \citet{Allen_2025} for star-forming galaxies at $3<z<4$ (median $z\simeq3.2$). We caution that this comparison relies on an extrapolation of their relation, as no galaxies in the \citet{Allen_2025} sample reach the stellar mass of ADF22.1. In both comparisons, ADF22.1 lies above the expected relations, i.e. it is significantly more extended than typical galaxies at comparable stellar mass and redshift. Thus, although one might qualitatively expect ADF22.1, being the brightest galaxy in the protocluster, to resemble a BGG, it does not follow the compact BGG trend at $z\sim3$. This offset could reflect a distinct evolutionary pathway driven by the dense protocluster environment, systematic differences in size measurement between samples, or a steepening of the size--mass relation at the high-mass end, analogous to what is observed $z = 0$ \citep{Shen_2003, Dutton_2011}. A larger, homogeneous analysis, incorporating structural parameters such as Sérsic index and adopting consistent fitting methodologies for field galaxies and galaxies in overdense regions, is required to disentangle genuine environmental effects from measurement systematics.

\subsection{The fate of ADF22.1}
Comparing the stellar mass and size of ADF22.1 with those of its potential early-type descendants offers useful insight into the mechanisms driving its evolution. Using the merger rates and mass-growth rates estimated by \citet{Conselice_2022}, \citet{Nipoti_2025} shows that minor and major dry mergers can plausibly increase stellar mass and size by factors of up to $\approx 2$--$3$ from $z = 3$ to $z = 0$. This is more than sufficient in the present case: the modest offsets in both stellar mass (a factor of 1.5) and size (a factor of 1.4; see Section~\ref{sec:masssize}) indicate that ADF22.1 is already broadly consistent with the $z = 0$ systems considered here, particularly given that the descendant candidates lie at the compact end of the local mass--size relation. Furthermore, comparing the baryonic mass of ADF22.1 with the median stellar mass of the descendant sample suggests that even moderate in-situ star formation prior to quenching could account for the required stellar mass growth, without invoking mergers.

As noted by \citet{Umehata_2025}, a key difference between ADF22.1 and local early-type galaxies concerns the stellar specific angular momentum. While $j_{\star}$ measurements are unavailable for MaNGA galaxies, some studies find that early-type galaxies follow a Fall relation with a lower zero-point than disc galaxies, regardless of whether they are fast or slow rotators \citep{Romanowsky_2012, Pulsoni_2023}. These results should nevertheless be treated with caution, as measuring $j_{\star}$ in the outskirts of early-type galaxies, where most of the specific angular momentum resides, is observationally challenging. Two evolutionary pathways can account for the properties of ADF22.1. In the first, ADF22.1 evolves into a descendant lying in the upper scatter of the early-type Fall relation, requiring no significant reduction in $j_{\star}$. In this case, the modest size and mass growth inferred from the comparison with the median descendant population (Section~\ref{sec:masssize}) suggests that internal dynamical processes \citep[e.g.,][]{Sellwood_2013}, perhaps combined with a few minor dry mergers, could be sufficient to transform ADF22.1 into an early-type galaxy. In the second pathway, if the descendant follows the early-type Fall relation of \citet{Pulsoni_2023}, $j_{\star}$ must decrease by a factor of $\sim$5 between $z = 3$ and $z = 0$. \citet{Umehata_2025} reached the same conclusion by comparing $j_{\star}$ of ADF22.1 with the early-type Fall relation of \citet{Romanowsky_2012}, invoking dry major mergers as the most plausible driver of this angular momentum loss \citep[see also][]{Zavala_2016, Lagos_2018a}. In this scenario, ADF22.1 would not evolve towards the median properties of the descendant sample, but rather towards the high-mass, large-size end of the distribution. Indeed, the most massive and extended galaxy in the descendant sample has a stellar mass and size a factor of 2.5 and 4 larger than ADF22.1, respectively. This pathway therefore implies substantial growth through dry minor and major mergers, which simultaneously increase stellar mass and size while reducing specific angular momentum, consistent with the scenario proposed by \citet{Umehata_2025}.

To summarize, regardless of the evolutionary pathway, it is likely that ADF22.1 is destined to become an extreme early-type galaxy: either with an unusually high-$j_{\star}$ or one of the most massive and extended galaxies in the local Universe.

\section{How do high-z giant discs form?}
\label{sec:formation}
The formation pathways typically invoked to explain the formation of giant discs at high redshift differ from those commonly associated with their counterparts at $z=0$. In the local Universe, giant discs are predominantly found in relatively isolated systems or low-density environments \citep[e.g.\ loose groups or cluster outskirts;][]{Ogle_2019}. In contrast, the two giant discs currently known at high redshift reside in overdense regions \citep{Wang_2025, Umehata_2025}. One widely discussed mechanism for assembling giant discs at high $z$ is sustained gas accretion with coherent angular momentum, for instance via filamentary inflows that preserve (or efficiently align) the spin of the incoming material \citep[e.g.,][]{Stewart_2013, Stewart_2017, Kret_2022}. An alternative pathway involves mergers of gas-rich galaxies on favourable orbits, where dissipation and angular-momentum redistribution can rebuild an extended, rotationally supported disc remnant \citep{Hopkins_2009, Lagos_2018a}. It remains unclear, however, whether these mechanisms, originally developed and often invoked to explain the high-$z$ disc population more generally, can operate efficiently in such extreme systems within overdense environments. Progress is currently limited by the fact that cosmological simulations with the spatial resolution required to robustly resolve disc structure at $z\gtrsim 3$ typically encompass volumes too small to capture the rarest, most massive galaxies in protocluster environments \citep{Jiang_2025}. In this section, we discuss the possible formation scenarios of ADF22.1 in light of the multiple diagnostics derived in the previous sections and assess their relative viability.

\subsection{Cold stream accretion}
Although cold-stream accretion from the cosmic web is one of the leading mechanisms proposed for the formation of giant discs \citep{Danovich_2015, Stewart_2013, Stewart_2017, Cadiou_2021}, the inflowing streams are expected to arrive along a wide range of trajectories as they approach the central galaxy \citep{Danovich_2015, Dekel_2020, Dekel_2020b, Hafen_2022}. This chaotic accretion can trigger violent disc instabilities, leading to highly turbulent and perturbed discs and promoting the formation of giant clumps \citep{Bournaud_2014, Krumholz_2018, Zolotov_2015, Dekel_2020b}. Internal torques then drive gas inflows and clump migration, ultimately building up a compact stellar component \citep{Zolotov_2015, Dekel_2020}. This framework struggles to account for the observed properties of ADF22.1 for several reasons. First, the virial temperature, estimated using the following relation, $T = 1.45 \times 10^ 6 (V_{\mathrm{c}}/200$ km s$^{-1})^2$ yields $T \sim 10^{7}$ K \citep{Cimatti_2019}, a regime in which coherent and global gas cooling is highly inefficient. One possible way to circumvent this limitation is the cold-in-hot accretion regime \citep[e.g.,][]{Dekel_2006, vandevoort_2012, Dubois_2012, Aung_2024, Waterval_2025}, in which, at $z>2$, cold gas filaments can still penetrate hot, massive haloes. However, even this scenario appears inconsistent with ADF22.1. Its high $V/\sigma$ ratio of $\gtrsim 6$ indicates a dynamically cold disc, unlike the discs expected from cold-stream accretion, which are typically highly turbulent, clump-dominated, and compact \citep[e.g.,][]{Bournaud_2014, Ginzburg_2022, Krumholz_2018}.

Using one galaxy at $z = 3$ from a zoom-in cosmological simulation, \citet{Kret_2022} showed that, under specific conditions, cold streams can promote the growth of dynamically cold discs, provided that the accreted material is largely co-rotating and approximately coplanar. They emphasize, however, that this phase is transient, lasting only a few hundred Myr. As their conclusions rest on a single case study, broader questions remain open: how common co-rotating, coplanar accretion events actually are, how frequently they give rise to dynamically cold discs, and how long such discs survive before being disrupted by misaligned cold gas accretion. Applying this scenario to ADF22.1 raises three main issues. First, the stellar component in the galaxy analysed in \citet{Kret_2022} is very compact, with an effective radius of 0.5 kpc compared to $\approx 5$ kpc for ADF22.1. This results in a halo-to-stellar size ratio of $R_{\rm 200}/R_{\mathrm{ e,m}}\sim 100$, a factor of 3 higher than observed in ADF22.1, $R_{\rm 200}/R_{\mathrm {e, m}}\sim 30$ (see Table \ref{tab:dyn_bestfit}). Second, the galaxy in \citet{Kret_2022} is dominated by a bulge component, with a bulge-to-total mass ratio of 0.7, whereas ADF22.1 is disc dominated, with a bulge-to-total mass ratio of 0.2. Third, in \citet{Kret_2022}, the dynamically cold phase is short-lived because subsequent accretion becomes increasingly counter-rotating and/or misaligned (including contributions from satellites), ultimately disrupting the gas disc. While a transient cold-disc phase in ADF22.1 cannot be excluded, the fact that both the gas and stellar discs are spatially extended argues against such a brief episode. The time required to assemble an extended stellar disc is typically many orbital periods because sustained star formation across radii and gradual angular-momentum redistribution are needed \citep{Sellwood_2002, Sellwood_2014, Vera_2014, Minchev_2026}. This is supported by a simple empirical comparison: the mean stellar age, which to order of magnitude goes as $M_{\star}/\mathrm{SFR} \sim 400$ Myr, exceeds the dynamical time (i.e., $2 \pi R_{\mathrm{e,m}}/V$) of $\sim 60$ Myr, by roughly an order of magnitude, implying that stars have had time to complete ${\sim}7$ orbits since they formed, consistent with the picture of slow, dynamically regulated disc growth.

To summarise, reproducing ADF22.1 through cold-stream accretion would require an unusually sustained and coherent inflow history. The accreted gas would need to remain largely co-rotating and co-planar for a sufficiently long period to assemble an extended, dynamically cold disc, while simultaneously avoiding the misaligned accretion events and satellite perturbations that simulations indicate would rapidly disrupt such an ordered configuration. In addition, a mechanism capable of preventing the compaction of the stellar component would need to be invoked. We therefore consider cold-stream accretion an unlikely primary explanation for the observed properties of ADF22.1.

\subsection{Giant discs hosted by haloes with hot CGM}
The fact that ADF22.1 resides in a dark-matter halo with a mass (and circular velocity) comparable to, or even larger than, those of massive discs in the local Universe suggests that some of the mechanisms invoked to explain local disc properties may already be in place at high redshift for galaxies living in overdense regions. 

One of the key mechanisms thought to be responsible for the formation and long-term survival of massive star-forming discs at low redshift is gradual cooling of a low-density and high temperature ($T\sim T_{\mathrm{vir}}$) circumgalactic medium (CGM), also called "corona" \citep{Armillotta_2016, Gronke_2020, Voit_2021, Afruni_2023}. In this picture, cool gas condense out of the hot CGM, lose buoyancy, and subsequently accrete onto the galaxy. In some cases, this mechanism can be stimulated and sustained by supernova (or perhaps also AGN) feedback in a fountain-like cycle, in which outflowing gas mixes with and accretes mass from the CGM before falling back onto the disc \citep{Fraternali_2017, Hobbs_2020, Barbani_2023}. This mechanism bears similarities to the condensation and precipitation processes observed in the cores of local galaxy clusters \citep{McDonald_2018}, as well as in protocluster environments at redshifts comparable to that of ADF22.1 \citep{Lepore_2024, Travascio_2025}. We therefore hypothesize that a related mechanism may contribute to the formation of ADF22.1. To our knowledge, no models or simulations currently quantify the impact of either spontaneous or fountain-driven hot gas condensation in very massive galaxy discs at high redshift. Nevertheless, below we outline how the key observed properties of our target could be accounted for within such a scenario:
\begin{itemize}
    \item Dynamically cold disc. Because this scenario proceeds in a relatively gentle manner, without major disruptive events, it does not destroy the gas disc. The result is a rotationally supported, dynamically cold configuration.

    \item Relatively high baryon-to-halo ratio. The observed baryon fraction $\epsilon_{\mathrm{b}}$ is roughly 43\% of the cosmic baryon fraction, a value comparable to that of $z = 0$ disc galaxies and higher than measured in local early-type galaxies with similar halo mass (see Section \ref{sec:SHMR}). Within a condensation-driven scenario, this could be interpreted as evidence for a relative inefficiency of the mechanisms (e.g., mergers and AGN feedback) that are typically invoked to prevent the hot corona from cooling and settling into the disc \citep{Moster_2018, Wang_Kai_2025}. 

\item High $f_j$, large disc size, and low Sérsic index. In the selective cooling scenario described in the Introduction, $f_{j_{\mathrm{b}}}$ is expected to be $< 1$. The measured value for ADF22.1 is instead $1.0_{-0.5}^{+0.7}$, which, while consistent with the standard selective cooling scenario within the uncertainties, warrants further discussion. In the case of spontaneous gas condensation, the high $f_{j_{\mathrm{b}}}$ and low Sérsic index can be interpreted within a modified selective cooling framework, in which only gas with specific angular momentum above a threshold ($j > j_{\min}$) is accreted onto the disc. This requires a mechanism to suppress accretion below $j_{\min}$; a plausible candidate is AGN feedback, which could prevent low-angular-momentum gas from reaching the disc. If, instead, low-angular-momentum gas is also accreted efficiently, then some form of angular momentum redistribution is required to explain the combination of a low Sérsic index and a flat rotation curve \citep{vandenbosch_2001b}. A galactic fountain offers a natural mechanism: by cycling gas through the CGM and reaccreting it at varying radii, it mixes material spanning a range of specific angular momenta, enhancing the fraction of gas at intermediate $j$ at the expense of both the low- and high-$j$ tails of the distribution. The large disc size follows naturally from the resulting high net specific angular momentum.
\end{itemize}



\subsection{The role of halo spin}
ADF22.1 has a stellar-to-halo specific angular-momentum ratio close to unity, albeit with large uncertainties. While this value is consistent with those measured for local massive discs \citep{DiTeodoro_2023, Romeo_2023, ManceraPina_2025b}, $f_{j_\star}$ may be overestimated if the true halo spin exceeds our assumed value of $\lambda = 0.035$, as we discuss below. 

First, relaxing the assumption on $\lambda$ and adopting $f_{j_\star} = 0.5$, as found in some simulations where $j_\star \sim j_\mathrm{DM}$ to within a factor of two 
\citep[e.g.,][]{Danovich_2015, Lagos_2017}, yields $\lambda \simeq 0.063$. The probability of drawing a halo with a spin in the interval $0.035 \lesssim \lambda \lesssim 0.063$ is $\sim 38\%$, indicating that such values are not uncommon. Second, since ADF22.1 resides in a pronounced overdensity, its halo spin may exceed our assumed value of $\lambda = 0.035$, as strong tidal torques in overdense environments can boost halo spin \citep{Ganeshaiah_2021}. In both cases, a higher halo spin would naturally account for the large disc size and high specific angular momentum of ADF22.1, given that classical disc formation theory predicts galaxy size to scale directly with halo spin \citep[e.g., $R_{\mathrm{e}} \propto \lambda R_{200}$;][]{Fall_1983, Mo_1998}. We caution, however, that the strength of this size--spin coupling varies 
significantly across galaxy formation models and remains an open question \citep{Somerville_2018, Jiang_2019, Liang_2025a, Liang_2025b}.

The potential link between the formation of giant discs and the high spin of the dark-matter halo is also consistent with recent results from \citet{Jiang_2025} who use the TNG100 simulation \citep{Springel_2018}, to search for analogues of the Big Wheel galaxy by selecting systems with low bulge fractions, stellar masses consistent with the observed system, and outliers in the mass–size relation. They identified five such galaxies, concluding that these discs represent only $\sim$1–2\% of galaxies at similar stellar masses. These systems form through a combination of factors: they experience fewer major mergers, accrete gas that is both more abundant and better aligned in angular momentum, and reside in haloes with systematically different properties compared to the parent population, namely higher spin parameters (by more than 20\%), lower concentrations, and more spherical shapes. However, we note that none of the simulated galaxies with sizes comparable to ADF22.1 reach the high $V/\sigma$ observed for ADF22.1. This possibly reflects the fact the TNG simulations does not include the modelling of the multiphase interstellar medium, so the resulting $V/\sigma$ values can be biased towards lower values \citep{Kohandel_2024, Semenov_2025}. Moreover, while our data do not allow us to place strong constraints on the merger history of ADF22.1, the observed lopsidedness may indicate recent minor mergers that did not destroy the disc. To be consistent with the presence of a well-defined stellar disc, any such interactions must have been sufficiently gentle to preserve the disc structure.

\section{Summary and conclusions}
\label{sec:conclusion}
Disc galaxies are expected to follow a mass–size relation that evolves strongly with redshift, with high-$z$ systems typically much more compact than local counterparts. However, recent \textit{JWST} observations have revealed two rare, unexpectedly large disc galaxies at $z \sim 3$ \citep{Wang_2025, Umehata_2025}, both located in overdense environments. In this paper, we focus on one of these systems, ADF22.1, and leverage the combination of high-resolution ALMA and \textit{JWST} observations to study its properties in detail. This work is enabled by the unusually extended [CII] emission of ADF22.1, which spans a diameter of $\sim30$ kpc. These data allow us to measure the rotation velocity and velocity dispersion, perform a rotation-curve decomposition, and constrain the mass of the dark-matter halo. 

We confirm that ADF22.1 reaches an exceptionally high outer rotation velocity of $\sim 530\,\mathrm{km\,s^{-1}}$, with a flat rotation curve extending from $R \approx 5\,\mathrm{kpc}$ out to $\approx 15\,\mathrm{kpc}$. Performing for the first time a rotation-curve decomposition for a galaxy of this type, we infer a halo mass of $\log (M_{200}/M_{\odot}) = 12.9^{+0.4}_{-0.3}$, a stellar-to-halo mass ratio of $0.2^{+0.3}_{-0.1}$, and a baryon-to-halo mass ratio of $0.4^{+0.6}_{-0.3}$ in units of the cosmological baryon fraction. We measure the specific angular momentum of the stellar, gaseous, and baryonic components, and further derive, from our dynamical model, stellar and baryonic-to-halo specific angular momentum ratios of $0.9^{+0.6}_{-0.4}$ and $1.0^{+0.7}_{-0.5}$, respectively. Taken together, these properties --- including the position of ADF22.1 in the stellar-to-halo and baryonic-to-halo mass relations, as well as in the stellar and baryonic Fall and Tully--Fisher relations --- indicate that ADF22.1 is structurally indistinguishable from local giant discs. The relatively high stellar-to-halo and baryon-to-halo mass ratios further suggest that AGN feedback, despite ADF22.1 hosting the brightest X-ray AGN in the protocluster, was insufficient at early epochs to expel enough gas to halt the growth of the disc.

We identify potential $z = 0$ descendants of ADF22.1 in the MaNGA survey by selecting galaxies with higher stellar masses and circular velocities than ADF22.1, and compare them to a mass-matched reference sample. Both samples occupy similar environments and halo masses, and are dominated by early-type galaxies. Notably, the descendant candidates predominantly lie on the compact end of the local mass--size relation, with sizes and stellar masses comparable to those already reached by ADF22.1 at $z \sim 3$. Two evolutionary pathways are possible: if the descendant retains relatively high $j_{\star}$, internal dynamical processes combined with minor dry mergers would suffice to complete the transformation; if instead it follows the average early-type Fall relation, $j_{\star}$ must decrease by a factor of $\sim 5$, requiring substantial dry major mergers that simultaneously increase stellar mass and size, pushing the descendant towards the most massive and extended outliers of the descendant sample.

Finally, we combine the extensive observational constraints on ADF22.1 to investigate its formation by comparing its properties with theoretical expectations. Cold-stream accretion, usually invoked to explain the formation of high-$z$ gas discs, is unlikely to account for this system: this mechanism typically produces turbulent, clumpy gas discs and compact, bulge-dominated stellar components, in clear tension with the stellar distribution and size, and the dynamically cold disc observed in ADF22.1. Instead, given that ADF22.1 appears dynamically mature and resides in a very massive halo, we consider an alternative pathway in which cold gas condenses out of its hot CGM, either spontaneously or through a fountain-like cycle driven by supernova (or AGN) feedback, naturally building a massive, extended, high-angular-momentum disc. Although dedicated simulations of this scenario at high redshift are not yet available, the proposed framework provides a physically consistent explanation for the key observed properties (i.e., dynamically cold disk, high baryon-to-halo ratio, large stellar size, high stellar-to-halo specific angular momentum ratio) of ADF22.1.

Crucially, this study has been possible only thanks to the unusually extended rotation curve of ADF22.1, which enables simultaneous constraints on both its baryonic and dark-matter components. Extending such measurements to larger samples of massive galaxies through deeper [CII] observations will be essential for fully understanding the interplay between galaxies and their host haloes in the early Universe. At the same time, determining whether similarly extended discs can form outside overdense environments, and whether such environments are a prerequisite for the formation of giant discs at high redshift, will require statistical studies of systems with well-characterised environmental properties.

\section*{Acknowledgements}

FR acknowledges support from the Dutch Research Council (NWO) through the Veni grant VI.Veni.222.146. PEMP is funded by NWO through the Veni grant VI.Veni.222.364.\\
ALMA is a partnership of ESO (representing its member states), NSF (USA) and NINS (Japan), together with NRC (Canada), NSC and ASIAA (Taiwan), and KASI (Republic of Korea), in cooperation with the Republic of Chile. The Joint ALMA Observatory is operated by ESO, AUI/NRAO and NAOJ. This work is based on observations made with the NASA/ESA/CSA James Webb Space Telescope. The data were obtained from the Mikulski Archive for Space Telescopes at the Space Telescope Science Institute, which is operated by the Association of Universities for Research in Astronomy, Inc., under NASA contract NAS 5-03127 for JWST. The JWST data underlying this article are publicly available and were accessed from the DAWN JWST Archive (DJA). DJA is an initiative of the Cosmic Dawn Center (DAWN), which is funded by the Danish National Research Foundation under grant DNRF140.

\section*{Data Availability}

The \textit{JWST} data underlying this article are publicly available and are associated with GO program \#3547. The ALMA data are publicly available; they are associated with the following programs ADS/JAO.ALMA\#2018.1.01306.S; ADS/JAO.ALMA\#2021.1.01406.S and were retrieved through the ALMA Archive website.



\bibliographystyle{mnras}
\bibliography{bib} 




\appendix
\section{Mass models}
\begin{figure*}
    \begin{center}  \includegraphics[width=\textwidth]{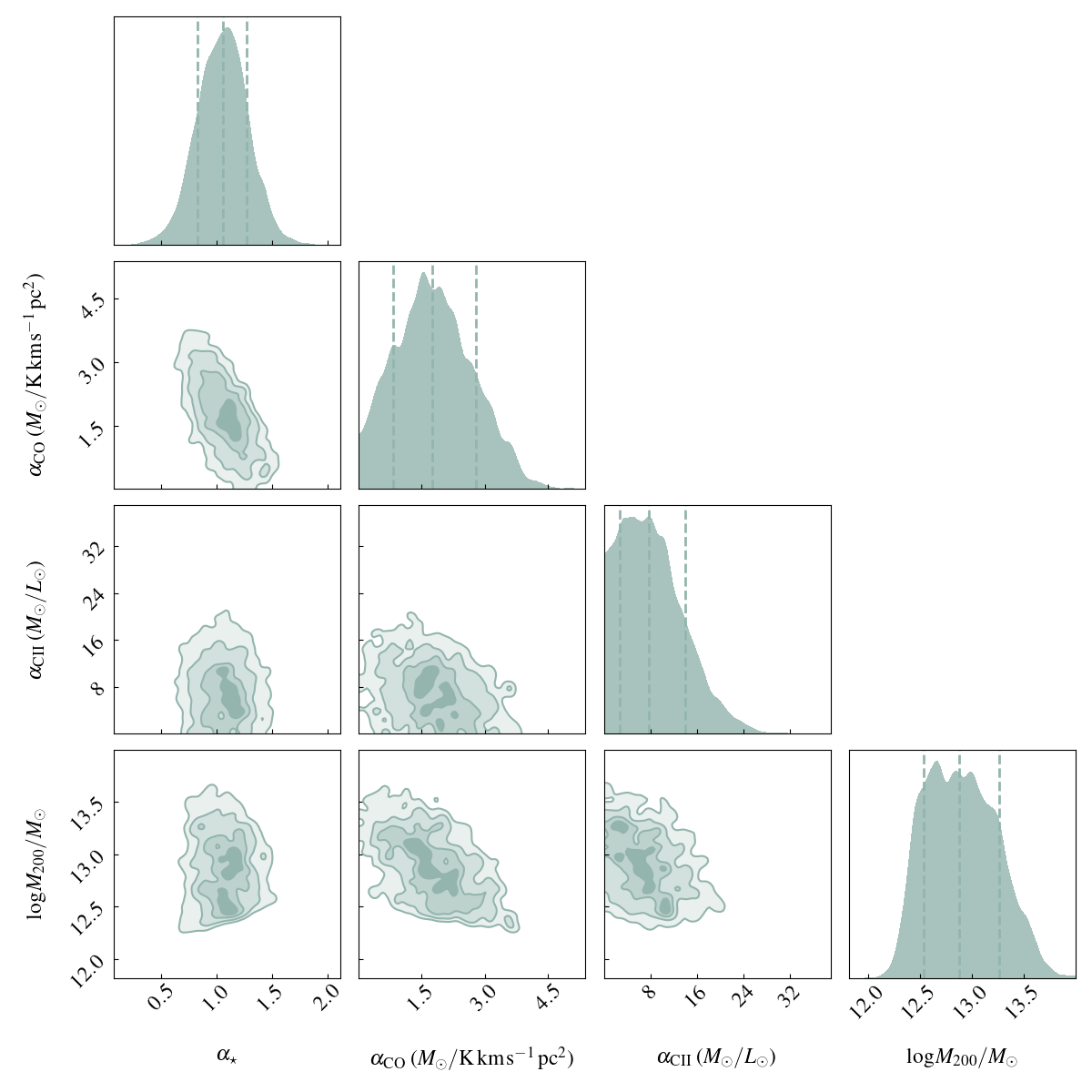}
        \caption{Posterior distribution for the parameters defining the rotation curve decomposition. The dashed lines correspond to the 16th, 50th and 84th percentiles.}  
        \label{fig:corner}
    \end{center}
\end{figure*}

\begin{figure*}
    \begin{center}  \includegraphics[width=\textwidth]{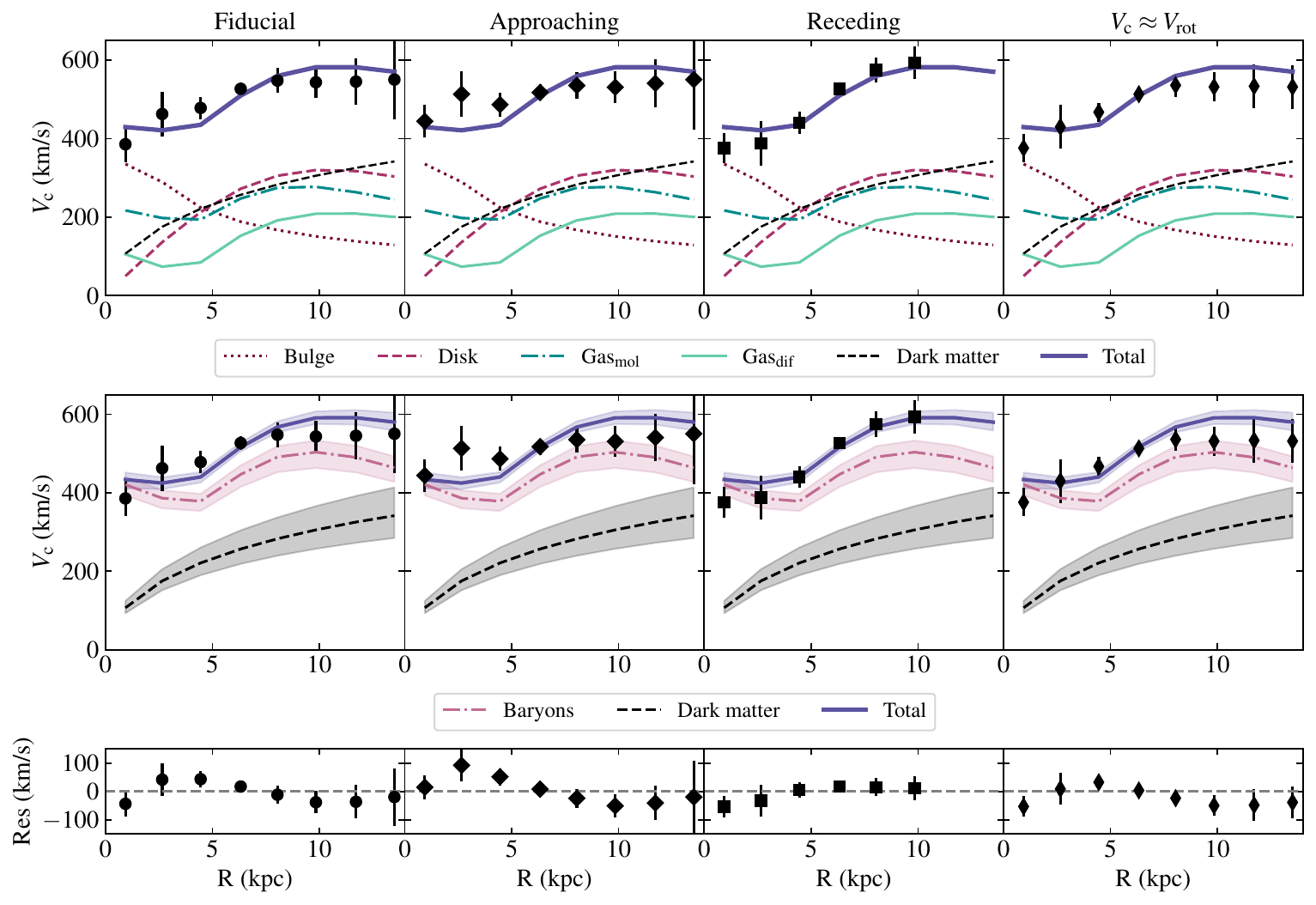}
        \caption{Circular velocity profiles derived under different assumptions, compared to the fiducial mass model. From left to right: the fiducial circular velocity curve (reproduced from Fig.~\ref{fig:decomposition_fiducial} for reference); the profile derived from the approaching side only; that from the receding side only; and the profile where the circular velocity is approximated by the rotation velocity. Top row: the individual baryonic and dark-matter components from the fiducial model. Middle row: the combined baryonic and total dark-matter profiles. Bottom row: residuals with respect to the fiducial model.}  
        \label{fig:decomposition_comp}
    \end{center}
\end{figure*}

In Fig.~\ref{fig:corner}, we show the posterior distributions of the best-fitting parameters for the fiducial rotation-curve decomposition shown in Fig.~\ref{fig:decomposition_fiducial}. In Fig.~\ref{fig:decomposition_comp}, we compare the mass models obtained from fitting the fiducial circular-velocity profile (left panels) with those derived from the circular-velocity profiles computed separately for the approaching and receding sides of the galaxy and for the circular velocity approximated by the rotation velocity (see Sect.~\ref{sec:vc} and Fig.~\ref{fig:vc}).

\bsp	
\label{lastpage}
\end{document}